\documentclass[sigconf]{acmart}
\usepackage{algorithm,algorithmic}
\usepackage{tabularx}
\usepackage{subcaption}

\AtBeginDocument{%
  \providecommand\BibTeX{{%
    \normalfont B\kern-0.5em{\scshape i\kern-0.25em b}\kern-0.8em\TeX}}}

\copyrightyear{2023}
\acmYear{2023}
\setcopyright{acmlicensed}
\acmConference[BuildSys '23]{The 10th ACM International Conference on Systems for Energy-Efficient Buildings, Cities, and Transportation}{November 15--16, 2023}{Istanbul, Turkey}
\acmBooktitle{The 10th ACM International Conference on Systems for Energy-Efficient Buildings, Cities, and Transportation (BuildSys '23), November 15--16, 2023, Istanbul, Turkey}
\acmPrice{15.00}
\acmDOI{10.1145/3600100.3623725}
\acmISBN{979-8-4007-0230-3/23/11}

\begin{document}

%\title{Scalable Multi-Channel Medium Access in Dynamic Wireless IoT Networks: A Multi-Armed Bandit Approach}

\title{PAMLR: A Passive-Active Multi-Armed Bandit-Based Solution for LoRa Channel Allocation}


\author{Jihoon Yun}
\affiliation{
  \institution{The Ohio State University}
  \city{Columbus}
  \state{Ohio}
  \country{USA}}
\email{yun.131@osu.edu}

\author{Chengzhang Li}
\affiliation{
  \institution{The Ohio State University}
  \city{Columbus}
  \state{Ohio}
  \country{USA}}
\email{li.13488@osu.edu}

\author{Anish Arora}
\affiliation{%
  \institution{The Ohio State University}
  \city{Columbus}
  \state{Ohio}
  \country{USA}}
\email{arora.9@osu.edu}
%\renewcommand{\shortauthors}{Yun, et al.}

\begin{abstract}
Achieving low duty cycle operation in low-power wireless networks in urban environments is complicated by the complex and variable dynamics of external interference and fading. We explore the use of reinforcement learning for achieving low power consumption for the task of optimal selection of channels. The learning relies on a hybrid of passive channel sampling for dealing with external interference and active channel sampling for dealing with fading. Our solution, Passive-Active Multi-armed bandit for LoRa (PAMLR, pronounced ``Pamela''), balances the two types of samples to achieve energy-efficient channel selection:  active channel measurements are tuned to an appropriately low level to update noise thresholds, and to compensate passive channel measurements are tuned to an appropriately high level for selecting the top-most channels from channel exploration using the noise thresholds.  The rates of both types of samples are adapted in response to channel dynamics. Based on extensive testing in multiple environments in different cities, we validate that PAMLR can maintain excellent communication quality, as demonstrated by a low SNR regret compared to the optimal channel allocation policy, while substantially minimizing the energy cost associated with channel measurements.

%LoRa is an emerging technology that enables long-range, low power consumption communication, particularly for urban environments. 
%One of the paramount challenges in LoRa communication is channel measurement, a critical factor in assisting LoRa devices to select optimal transmission channels. 
%Active channel measurements, although accurate, require transmitting pilot signals, a process that is notably energy-intensive and not ideal for cost-effective LoRa devices. 
%To address this, we introduce a solution termed Passive-Active Multi-armed Bandit for LoRa (PAMLR, pronounced ``Pamela'') in this paper.
%PAMLR adopts a balanced approach by incorporating both passive and active channel measurements to achieve energy-efficient channel allocation for LoRa networks.
%The main function of PAMLR is to utilize a small portion of active channel measurements to update noise thresholds, while leveraging a larger portion of passive channel measurements for selecting the top-$k$ channels from channel exploration using the noise thresholds. %{\color{red} please write one  sentence.}
%To evaluate the effectiveness of PAMLR, we conducted extensive tests in multiple environments in different cities
%%Columbus, Ohio, and Brooklyn, New York.
%The results attest that PAMLR can maintain excellent communication quality, demonstrated by a low SNR regret compared to the optimal channel allocation policy, while substantially minimizing the energy cost associated with channel measurements.

\end{abstract}

%\ccsdesc[500]{Computing methodologies~Neural networks}

%\keywords{Low-power, Robustness, Embeddings, LoRa, External interference, Fading, Infrastructure-free, Smart cities}


\maketitle

\section{introduction}
Wireless communication technologies have rapidly evolved over the last few decades, leading to increased adoption of robust, long-range, and energy-efficient solutions for a variety of applications. A particular class of networks that has seen widespread adoption due to its unique characteristics is the Low Power Wide Area Networks (LPWAN) \cite{LPWAN_Sigfox, LPWAN_NB_IoT, LPWAN_HaLow, LPWAN_DASH7}. Within this group, the Long Range (LoRa) \cite{LPWAN_LoRa} technology has emerged as a significant player. LoRa is a radio modulation technique for LPWAN used in the Internet of Things (IoT) and machine-to-machine (M2M) applications. Its combination of long-range capabilities and low power consumption makes it an ideal choice for urban environments where device density can be high, and the energy constraint is a critical factor.

Attaining low duty cycle operation in urban LoRa networks presents multifaceted considerations. The density and diversity of the urban environment introduces unpredictable dynamics in external interference and fading. As these factors can significantly impact the reliability of communications, continual selection of optimal channels becomes key, which in turn requires potentially frequent precise channel measurements. Yet, the energy cost associated to these measurements must also be kept minimal, in order to satisfy the stringent energy constraints that are often associated with LoRa networks.

When it comes to channel selection, active measurement techniques, as demonstrated by \cite{active_ch_1, active_ch_2, active_ch_3, active_ch_4, active_ch_5, active_ch_6, active_ch_7, active_ch_8}, involve the use of pilot signals from the transmitter, such as the Signal-to-Noise Ratio (SNR), Received Signal Strength Indicator (RSSI), and Packet Reception Rate (PRR), notably, on all channels. This approach provides accurate channel information, even if it conflates the impacts of external interference and fading and is energy-intensive due to the pilot signal transmissions. On the other hand, passive measurements, as employed by \cite{passive_ch_1, passive_ch_2, passive_ch_3, passive_ch_4, passive_ch_5}, which are obtained by simply listening and acquiring interference and noise values, are substantially more energy-efficient. But they offer only limited channel information, in particular eschewing information related to fading.  Clearly, there is a discernible trade-off for the two types of measurements.  Thus, a judicious design of how to combine and schedule for these two types of measurements is essential. While the matter has received considerable study, we find that there is substantial room for improvement.
%{\color{red} Please write a paragraph to summarize existing work on channel measurements (please include active and passive measurements. Please include some references (as many as possible), and state their limitations on this LoRa setting. }
%Active measurements can deliver a more precise channel information, albeit at the expense of transmitting pilot signals, making it more energy-intensive. 
%Conversely, passive measurements are more energy-efficient, though they can only yield limited channel information.

With that in mind, in this paper, we propose a novel reinforcement learning-based approach for optimal channel selection that aims to minimize power consumption without compromising communication reliability.  The solution, which we call Passive-Active Multi-Armed Bandit for LoRa (PAMLR, pronounced ``Pamela''), combines passive and active channel sampling in a unique way.  

The central concept for channel selection in PAMLR is a noise threshold. Its design lets PAMLR predict whether the current noise from passive channel measurements will impact the communication's reliability for the current fading condition. PAMLR assigns each frequency its own noise threshold, which is generated using RSSI values obtained from active channel measurements. 
%{\color{red} Please write one paragraph to describe the noise threshold and how it works in PAMLR.}

PAMLR balances the active and the passive sampling strategies to achieve energy-efficient channel selection as follows. Given the presence of fading, active measurements across all channels are needed in principle.  Rather than measure actively across all channels in an ongoing fashion as existing works do, which is particularly unfavorable for low-cost LoRa devices \cite{LoRa_Power}, PAMLR limits itself at any time to active measurement of only the current top-most channels in terms of reliability.  These measurements leads to updating the noise threshold for each of these channels.

In addition to calibrating the impact of fading (along with interference) in terms of the noise threshold, PAMLR concurrently performs estimation of the reliability of each channel.  More precisely, based on the noise thresholds, PAMLR employs a Multi-Armed Bandit (MAB) \cite{MAB} algorithm specifically designed for efficient channel exploration with passive samples only. The channel exploration selects the top-most channels based on the rewards obtained from comparing the total number of noise values from passive channel measurements that are higher than or equal to their noise thresholds, as well as those that are lower than the thresholds. While passive measurement in relative terms to active measurement has very low energy cost, this exploration is designed to start off with exploring all frequencies randomly but to progressively explore a diminishing number of frequencies over time.

PAMLR tunes active channel measurements to a suitably low level to update noise thresholds of the top-most channels. Concurrently, it tunes passive channel measurements to a suitably higher level, allowing for update of the top-most channels.  It also performs this compensation of lower active measurements with higher passive measurements, within limits, adaptively, to deal with channel dynamics.

We summarize the main contributions of this paper as follows:
\vspace*{.25mm}
\begin{itemize}
 %   \item We highlight the critical challenge of channel measurement in LoRa communication, especially concerning the energy-intensive nature of active channel measurements which can be a significant drawback for low-cost LoRa devices. 
    \item We propose a novel multi-armed bandit solution for LoRa networks that ---within limits--- allows for compensating active channel measurements with  passive channel measurements. 
    
    \vspace*{1mm}
    \item Compared to traditional approaches relying partly or solely on active measurements, PAMLR achieves substantial reduction in energy consumption. It outperforms state-of-the-art learning-free methods, i.e.,  \cite{active_ch_5}, by one to two orders of magnitude and state-of-the-art learning-based methods, i.e., \cite{active_ch_7}, by one order of magnitude in terms of energy consumption.
   % \item To achieve the design objectives of PAMLR, we employ the following methodologies. The noise thresholds are updated through active channel measurements. The channel exploration process involves passive channel measurements using the noise threshold, and the top-most channels are subsequently selected. 
    %... {\color{red} Please write one paragraph to present some detailed design of PAMLR, including noise threshold and channel exploration.}
    
    \vspace*{1mm}
    \item We provide comprehensive validation of PAMLR using extensive testing in diverse environments across several cities. Our results illustrate that PAMLR not only maintains superior communication quality---with a low SNR regret relative to an optimal channel allocation policy---but also curtails energy expenditure associated with channel measurements. Moreover, we empirically demonstrate how increasing passive measurements compensates for smaller active measurements, achieving similar SNR regret with reduced energy consumption. \\ \ \\
\end{itemize}

\vspace{-1.1cm}
\section{System Model and Problem Statement}
\label{sec:model}

In LoRa networks, source nodes typically generate environmental or  other data, which they propagate to one or more collector nodes.  If the network is a mesh, the propagation can be via adjacent nodes. To maximize energy efficiency, source nodes need to operate on a low enough duty cycle that suffices for the computing and communication needs, with the rest of their time spent in a dormant state. (To further reduce their duty cycle, nodes may optionally synchronize their wake-up time with specified time slots allocated for their respective tasks, which allows their neighbors to learn when the nodes are active.)

Consider for simplicity the scenario involving one transmitter node and one receiver node in a LoRa network\footnote{The results of this paper are readily extended for the set of links in a star or a multi-hop mesh network.}.
%Adhering to a receiver-centric MAC protocol, such as RC-MAC \cite{huang2014rc} or O-MAC \cite{omac}, the receiver node  periodically wakes up and signals the transmitter node to initiate a communication cycle. 
%During this phase, 
The channels between the two may be subject to external interference, which may vary over time and be frequency specific. Also, the transmissions may be subject to fading as a result of obstacles, and may also vary over time and be frequency specific. To address these impediments, both nodes need to identify common frequency bands that will be used for their communications. Assume that there are $N$ total available frequency bands and the nodes needs to select $k$ ($k\leq N$) of these for transmission. The primary focus of this paper is the energy efficiency of the method for the selection of these $k$ frequency bands subject to maximizing transmission reliability.

\begin{table}[ht] 
\centering
\caption{Notations}    \label{tab:notation}
\setlength{\tabcolsep}{1pt}
\renewcommand{\arraystretch}{1.1}
\begin{tabularx}{\linewidth}{lX}

\hline 
 Symbol  &  Definition  \\ \hline
$N$ & The total number of channels available\\
$k$ & The number of channels selected for utilization\\
$\Omega$ & The discount factor for noise rewards\\
$\omega$ & The weighting factor for EWMA for RSSI and noise \newline values\\
$\rho^{i}_{r}(t)$  & The current RSSI value at channel $i$ and time $t$  \\ 
$\rho^{i}_{n}(t)$  & The current noise value at channel $i$ and time $t$  \\ 
$\rho^{i}_{EW\!MA,r}(t)$  & The EWMA of RSSI values at channel $i$ and time $t$  \\ 
$\rho^{i}_{EW\!MA,n}(t)$  & The EWMA of noise values at channel $i$ and time $t$  \\ 
$\rho^{i}_{th,n}(t)$  & The threshold of noise values at channel $i$ and time $t$  \\ 
$\gamma^{min}_{SF}$  & The minimum SNR based on LoRa Spreading Factor (SF)  \\ 
$\alpha^{i}(t)$ & The total number of noise values lower than the noise \newline threshold at channel $i$ and time $t$\\
$\beta^{i}(t)$ &  The total number of noise values higher than or equal \newline to  the noise threshold at channel $i$ and time $t$\\
$\theta^{i}(t)$ & The reward, derived from a Beta distribution with \newline parameters $\alpha$ and $\beta$\\
$\gamma^{i}(t)$  & The SNR value at channel $i$ and time $t$\\ 
$C^{k}_{alg}$ & A set of k channels selected by the algorithm\\
$C^{k}_{opt}$ & A set of k optimal channels \\
%$\gamma^{max,k}_{avg,alg}$  & The average of the maximum SNR values of k number selected by the algorithm  \\ 
$E_{a}$ & The energy expenditure for active information\\ 
$N_{a}$ & The number of active measurements for a time slot \\
$E_{p}$ & The energy expenditure for passive information \\ 
$N_{p}$ & The number of passive measurements for a time slot \\
$E$ & The total energy expenditure for network quality \newline information \\
$R_{SNR}$ & The regret for SNR\\

\hline 
\end{tabularx}
\end{table}

%Table~\ref{tab:notation} presents the key notations used in this paper. 

%In traditional multi-frequency MAC protocols, the primary focus has been on resolving collisions caused by internal interference to optimize throughput. However, in real-world scenarios, the presence of unexpected external interference and fading can significantly impact reliability, especially in large-scale environments like cities. To tackle these challenges, it becomes essential to accurately assess the network quality of assigned frequencies using the protocol.

%In this paper, we propose a multi-frequency medium access approach that integrates a well-balanced combination of active and passive information, with a particular emphasis on passive information. Our approach aims to achieve higher energy efficiency, a fundamental requirement for IoT devices, while ensuring the reliability of the communication. By leveraging both active and passive information, our system strives to enhance network quality performance.

%Moreover, our system ensures sustained high reliability over time by dynamically updating the set of frequencies. This flexibility enables the system to adapt to temporal variations, resulting in improved overall performance, responsiveness, and a consistently reliable connection.

Table~\ref{tab:notation} summarizes the notations we use henceforth.

More formally, let's say that before each communication cycle $t$, the receiver can take $N_p(t)$ passive measurements and $N_a(t)$ active measurements among the $N$ available frequency bands.
Let $\lambda_p$ and $\lambda_a$ denote the average number of passive and active measurements taken per communication cycle, respectively.
Then we have 
\begin{equation}
    \lambda_p=\lim_{T\to\infty} \frac{1}{T}\sum_{t=1}^TN_p(t),
\end{equation}
and
\begin{equation}
    \lambda_a=\lim_{T\to\infty} \frac{1}{T}\sum_{t=1}^TN_a(t).
\end{equation}

One objective of this paper is the energy cost for channel measurements at the receiver.
Let $E$ denote the average energy cost  for channel measurements per communication cycle, which can be computed as the sum of the energy used for both passive and active measurements.
Let $E_p$ and $E_a$ denote the energy costs for each passive and active measurement, respectively.
Then $E$ is given by
\begin{equation}
E = \lambda_pE_p+\lambda_aE_a.
\end{equation}

In LoRa, it is noteworthy that $E_p$ typically falls several orders of magnitude below $E_a$. For instance,  let's consider a data rate in the tens of kilobits per second (kbps) and a communication range of several hundred meters in an urban environment. Achieving this goal relies on specific configuration parameters, including a Spreading Factor (SF) of 8 and a 500 KHz bandwidth. 
%These parameters play a pivotal role in determining both the data rate and communication range while also directly impacting power consumption.
Additionally, we consider a transmission power of 20 dBm and a 5-byte packet payload. In this context, $E_{a}$ and $E_{p}$ represent 23.89 mW and 0.023 mW, respectively, for active and passive information \cite{sx1276}. This means $E_{a}$ is approximately 1038 times larger than $E_{p}$
Consequently, in the design of our algorithm, the primary strategy for minimizing $E$ hinges on minimizing the rate of active measurements, while permitting a relatively high rate of passive measurements.

%For instance, with a specific configuration parameterization such as a Spreading Factor (SF) of 8, Bandwidth of 500 KHz, transmission power of 20 dBm, and a packet payload of 5 bytes, $E_{a}$ and $E_{p}$ stand at 23.89 mW and 0.023 mW respectively for active and passive information \cite{sx1276}. This means $E_{a}$ is approximately 1038 times larger than $E_{p}$.

%In the case of LoRa, when using specific configuration parameters such as Spreading Factor (SF) 8, Bandwidth 500 KHz, transmission power 20 dBm, and packet payload 5 bytes, the calculated values for energy expenditure are $E_{a}$ = 23.89 mW for active information and $E_{p}$ = 0.023 mW for passive information. These values highlight a significant difference in energy consumption between active and passive information acquisition in LoRa networks, with $E_{a}$ being approximately 1038 times larger than $E_{p}$ \cite{sx1276}.

Alongside minimizing energy cost $E$, another objective of this paper is to improve transmission reliability, that is, to maximize the PRR. Nonetheless, it is challenging to predict PRR based solely on channel information such as SNR, RSSI, and the like.
Consequently, this paper will deviate from directly using PRR as our objective and instead adopt the SNR all communication cycles as the objective. 
Given that PRR generally increases with SNR, we can posit that maximizing SNR will typically result in a maximized PRR.
In particular, let $R_\text{SNR}(T)$ denote the regret of SNR, which quantifies the discrepancy between the SNR of a set of $k$ channels selected by the algorithm during the time period from $t=T-W+1$ to $t=T$, and the SNR of an optimal set of $k$ channels during the same time period. 
Deviating from the conventional definition of regret in reinforcement learning, which typically encompasses the period from $t=1$ to $t=T$, our algorithm's evaluation centers around the time window from $t=T-W+1$ to $t=T$. This choice is driven by the continuous fluctuations in SNR over time, allowing for a more accurate assessment of our algorithm's ability to adapt to varying channel conditions.

Let $\gamma_i(t)$ denote the SNR values for channels $i$ and at time $t$.
Let $\mathcal{C}_\text{alg}^t$ and $\mathcal{C}_\text{opt}^t$ denote the set of the $k$ channels selected by the algorithm at $t$ and the set of the $k$ optimal channels (the $k$ channels with highest SNR) at time $t$, respectively.
Then  $R_\text{SNR}(T)$ is given by
\begin{equation}
R_\text{SNR}(T)= \frac{1}{W}\sum\limits_{t=T-W+1}^{t=T}\big(\sum\limits_{i \in \mathcal{C}^{t}_\text{opt}} \gamma_{i}(t) \ - \sum\limits_{ j \in \mathcal{C}^{t}_\text{alg}} \gamma_{j}(t)\ \big).
\end{equation} 

We can now formally define our problem as follows.
Assuming the rates and energy costs of passive and active measurements, i.e., $\lambda_p$, $\lambda_a$, $E_p$, and $E_a$, are all given, our goal is to find an optimal channel measurement strategy to minimize the SNR regret $R_\text{SNR}(T)$. 
In particular, this involves determining how to select $N_p(t)$ frequency bands for passive measurement and $N_a(t)$ bands for active measurement at each time instance $t$, and how to select the set $\mathcal{C}_\text{alg}^t$ based on these measurements, with the goal of minimizing $R_\text{SNR}(T)$.

%\section{Channel Selection based on Channel Exploration and Noise Threshold}
\section{PAMLR: A Multi-Armed Bandit-Based Solution}
\begin{figure}[t]
	\centering
	\includegraphics[width=0.27\textwidth, trim = 0 0 0 0, angle=-90]{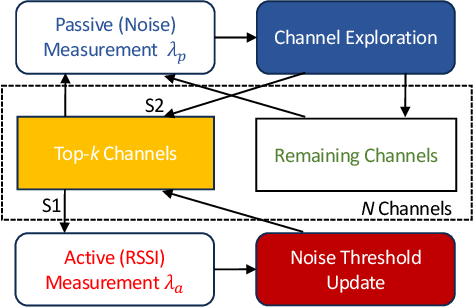}
	\caption{PAMLR system design}
	\label{fig:diag}
\end{figure}

In this section, we describe PAMLR, our solution to the LoRa scheduling problem detailed in Section \ref{sec:model}. Figure \ref{fig:diag} illustrates the overall design of PAMLR. Recall that at the beginning of each time slot $t$, the transmitter needs to select $k$ channels from $N$ channels for transmission, i.e., the yellow block in Figure \ref{fig:diag}. This selection can be achieved in two ways.  The first method utilizes passive measurements, depicted by the two upper blocks in Figure \ref{fig:diag}. The second method involves active measurements, illustrated by the two lower blocks in Figure \ref{fig:diag}. Each of these methods are detailed separately below.

\begin{figure}[t]
	\centering
	\includegraphics[width=0.35\linewidth, trim = 8 0 0 0, angle=90]{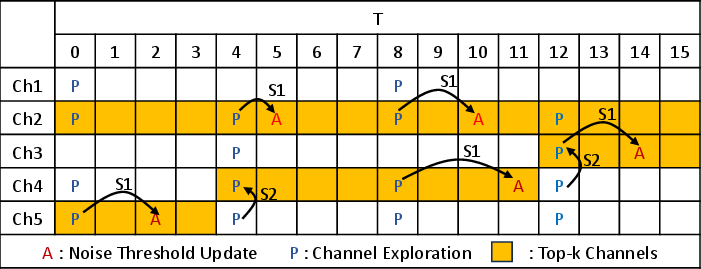}
	\caption{Illustration of channel allocation:  rare active measurements are balanced with frequent passive measurements}
	\label{fig:illustration}
\end{figure}

%{\color{red} Please reorganize the following paragraphs into two paragraphs to briefly discuss passive measurements and active measurements.}

PAMLR incorporates a channel exploration process that begins with passive measurements at the rate of $\lambda_{p}$ to measure the background noise. These measurements are used to identify a set of top-$k$ channels from all $N$ channels based on the noise threshold.

The selected top-$k$ channels undergo active measurements (RSSI) at the rate of $\lambda_{a}$ to update the noise threshold. The noise threshold serves as a criterion to evaluate whether the background noise, recorded during passive measurements, is below or above the defined threshold level and accordingly yields a metric denoted as $\theta^{i}(t)$ for channel $i$ and each time $t$.  
%This metric is derived from a Beta distribution, utilizing the count of noise values below the threshold and those equal to or exceeding the threshold. 
PAMLR leverages this metric to make channel selection decisions for transmission.
 %which is used by PAMLR to select channels for transmission.

\noindent {\bf Illustrative example.\/} \ \ \
By way of an illustration, Figure~\ref{fig:illustration} presents an example of PAMLR spanning a duration of $T=15$ time units, with $N=5$ and $k=2$ channels. Additionally, the values of $\lambda_{p}$ and $\lambda_{a}$ are respectively set to 1 and 0.2. Initially, i.e., at time $t=0$, none of the top-$k$ channels are  selected as such. As PAMLR conducts random passive measurements and channel exploration on a set of 2$k$ channels, as shown in Figure~\ref{fig:illustration}, PAMLR selects Ch1, Ch2, Ch4, and Ch5 at $t=0$ for passive measurement and channel exploration.

The channel exploration selects Ch2 and Ch5 as top-$k$ channels at $t=0$, due to their higher $\theta^{i}(t)$ compared to others. At $t=3$, Ch2 performs active measurement and updates the noise threshold, denoted by action S1, and Ch5 performs likewise at $t=6$. Since active measurements require communication between nodes, Ch2 and Ch5 cannot perform them simultaneously. The timing of these measurements and noise threshold updates is determined by $\lambda_a$, ensuring they occur at different times.

At $t=4$, passive measurements and channel exploration are conducted for 2$k$ channels, which include the top-$k$ channels, Ch2 and Ch5, as well as $k$ other channels, Ch3 and Ch6, selected based on the highest $\theta^{i}(t)$s excluding Ch2 and Ch5.  This results in a change in the top-$k$ channel selection, shifting from Ch5 to Ch4. Ch4 demonstrates higher $\theta^{i}(t)$ compared to Ch5, indicated by action S2. Meanwhile, Ch2 maintains its position as one of the top-$k$ channels due to its consistently high $\theta^{i}(t)$.

Similarly, at $t=9$ and $t=13$, passive measurements and channel exploration are conducted, and the top-$k$ channels are changed based on the highest $\theta^{i}(t)$. The process of exploring and selecting the top-$k$ channels continues throughout the duration of $t=15$, ensuring the adaptation and optimization of channel selection in PAMLR.
\hfill
\qed

%Firstly, active measurements exclusively focus on the top-$k$ channels, denoted as action S1. {\color{red} Please explain with more details for the first several time slots. E.g., at $t=1$, channel exploration are taken in ch1, ch2, ch4, and ch5, at $t=2$, ... }
%Following channel exploration, the set of top-$k$ channels can either be updated (action S2) or remain unchanged. Secondly, active measurements occur infrequently, while passive measurements are conducted more frequently. Thirdly, the passive measurement phase involves exploring additional channels beyond the top-$k$ set. In this example, 2$k$ channels are examined to identify potential alternatives if the current top-$k$ channels prove to be incorrect or exhibit performance degradation. Specifically, at $t$=9 and $t$=13, passive (noise) measurements and channel exploration take place, resulting in a change from $Ch$6 to $Ch$3 at $t$=13.

%{\color{red} I can't understand this example very well. Do you think it will be a better idea to present this example later, e.g., somewhere in section 3.2?}
 
The following two subsections provide the two critical components of PAMLR in detail: Noise Threshold Update based on active measurements and Channel Exploration based on passive measurements.

%\begin{figure}[t]
    %\vspace{1mm}
%	\centering
%	\includegraphics[width=0.9\linewidth]{images/diag.pdf}
 %   \vspace{-6mm}
%	\caption{The }
%	\label{fig:diag}
%	%\vspace{-5.5mm}
%\end{figure}

%\begin{figure}[t]
    %\vspace{1mm}
%	\centering
%	\includegraphics[width=1\linewidth]{images/arch.pdf}
%    \vspace{-6mm}
%	\caption{The }
%	\label{fig:arch}
	%\vspace{-5.5mm}
%\end{figure}

\subsection{Noise Threshold Update}

\emph{Noise threshold} is a central concept in the design of PAMLR, cf.~the bottom-right block in Figure \ref{fig:illustration}. For each channel $i=1,2,\cdots,N$, the transmitter maintains a noise threshold, denoted by $\rho^{i}_{th,n}(t)$ for channel $i$ and each time $t$. The noise threshold is used to predict whether the noise derived from passive measurements will have an impact on the reliability of communication.  
%Now we look at its definition.

\subsubsection{Definition of Noise Threshold for Stationary Channels}
We first consider a simplified scenario where the channels are stationary, i.e., they don't exhibit dynamics. Note that in the passive-active framework of LoRa presented in Section \ref{sec:model}, the information obtained from passive measurements is limited to being primarily associated with external interference and channel noise, rather than signal fading.  To assess the potential impact of the interference and noise on the communication, it suffices to have a metric based on passive measurement only; we design this metric in terms of the noise threshold.

The definition of noise threshold is based on the following considerations. For LoRa devices, different spreading factors (SFs) have different minimum SNR requirements for a transmission to be successful, denoted by $\gamma^{min}_{SF}$ and restated in Table~\ref{tab:min_SNR} (source: \cite{sx1276}). In other words, for a specific SF, LoRa can only successfully decode the received signal if the SNR surpasses its minimum threshold.  Typically, the minimum threshold for LoRa devices is negative, since LoRa devices are designed to receive packets even when the noise level surpasses the signal power.

Taking these considerations into account and based on the relationship between SNR, RSSI, and noise, we define the noise threshold of channel $i$, denoted by $\rho_{th,n}^{i}$, as follows:
\begin{equation}\label{def:noise_threshold}
\rho_{th,n}^{i} = \rho_{r}^{i} - \gamma_{SF}^{min},
\end{equation}
where $\rho_{r}^{i}$ represents the RSSI value obtained through active measurement at channel $i$, while $\gamma_{SF}^{min}$ denotes the minimum SNR based on the user-set SF in the LoRa configuration. 

%{\color{red} What is RSSI? How do we obtain it? And, what is $\gamma_{SF}^{min}$ and how do obtain  $\gamma_{SF}^{min}$? Please write a few sentences (pharhps some equations) to define these clearly. Please note that whenever you bring out a new notation, you need to define it clearly, without any ambiguity.} 

\begin{table}
  \caption{Minimum SNR requirement for various LoRa Spreading Factors (SF)}
  \label{tab:min_SNR}
  \begin{tabular}{cccccccc}
    \toprule
    SF& 6 & 7 & 8 & 9 & 10 & 11 & 12\\
    \midrule
    $\gamma_{SF}^{min}(dB)$ & -5 & -7.5 & -10 & -12.5 & -15 & -17.5 & -20 \\
  \bottomrule
\end{tabular}
\end{table}

The intuition for this definition is as follows. 
%SNR is a fundamental metric to assess the quality of a signal in relation to the noise present in the system. It represents the ratio of the received signal power to the noise power. 
In the case of LoRa, if the difference between the current RSSI and the current noise level exceeds the minimum SNR requirements, it follows that the signal can be successfully demodulated by the LoRa receiver. Therefore, the noise threshold as defined serves as a reference for assessing the signal reliability.

%{\color{red} Please write one paragraph to elaborate the ideas.}

\begin{comment}
we have defined the noise threshold $\rho_{th,n}^{i}$, which represents the current threshold for noise values at channel $i$. 
The calculation of the noise threshold $\rho_{th,n}^{i}$ involves two key parameters: $\rho_{r}^{i}$, denoting the current RSSI value at channel $i$, and $\gamma_{SF}^{min}$, representing the minimum SNR based on the LoRa SF.
{\color{red} Please elaborate this paragraph. This defination}

\begin{equation}
\rho_{th,n}^{i} = \rho_{r}^{i} - \gamma_{SF}^{min}.
\end{equation}
\end{comment}

\subsubsection{Adapting the Noise Threshold for Non-Stationary Channels}
The channel in the general case is dynamic and, especially in urban conditions, can change frequently. To keep the noise threshold current, we employ the exponentially weighted moving average (EWMA) of RSSI, denoted as $\rho^{i}_{EW\!MA,r}$. The weighting factor $\omega$ in EWMA allows us to prioritize either the current value or the historical data by choosing different values for $\omega$. The choice of $\omega$ depends on the environmental conditions. When environmental conditions are constantly changing, selecting $\omega$ close to 1 enables faster adaptation to these changes. Conversely, in stable environmental conditions, a smaller $\omega$ is preferable to reduce the impact of outliers. The expression for $\rho^{i}_{EW\!MA,r}(t)$ is given by:
%\vspace*{-.5mm}
\begin{equation}
\rho^{i}_{EW\!MA,r}(t) = \omega \cdot \rho^{i}_{r}(t) + (1-\omega) \cdot \rho^{i}_{EW\!MA,r}(t-1).
\end{equation}

\noindent Based on the EWMA of RSSI, the noise threshold $\rho^{i}_{th,n}(t)$ becomes:
%\vspace*{-.5mm}
\begin{equation}
\rho^{i}_{th,n}(t) = \rho^{i}_{EW\!MA,r}(t) - \gamma^{min}_{SF}.
\end{equation}

\subsubsection{Incorporating Packet Loss}
Packet loss is key to estimating channel quality. However, when packet loss occurs, it is impossible to obtain the corresponding RSSI and SNR measurements. To accommodate the impact of loss, we employ an alternative approach: we utilize passive information, specifically noise measurements. Unlike RSSI and SNR, noise can be acquired regardless of packet loss. In keeping with LoRa Radio specifications, we assume that the reported RSSI value measures the current noise added to the minimum SNR based on the LoRa SF. To account for non-stationary channels, we utilize the EWMA of noise. The expression for $\rho^{i}_{EW\!MA,r}(t)$ when packet loss occurs is given by:
%\vspace*{-.5mm}
\begin{equation}
\rho^{i}_{EW\!MA,r}(t)=\omega\cdot(\rho^{i}_{EW\!MA,n}(t)+\gamma^{min}_{SF})+(1-\omega)\cdot\rho^{i}_{EW\!MA,r}(t-1)
\end{equation}

The complete algorithm for Noise Threshold Update in PAMLR is given in Algorithm \ref{alg:NT}.

\begin{algorithm}[t]
\caption{Algorithm for Noise Threshold Update}\label{alg:NT}.
\begin{algorithmic}[1]
\renewcommand{\algorithmicrequire}{\textbf{Input:} }
\renewcommand{\algorithmicensure}{\textbf{Output:}}
\REQUIRE $\rho^{i}_{r}(t), \rho^{i}_{n}(t)$, $\alpha^{i}(t-1)$, $\beta^{i}(t-1)$, $\theta^{i}(t-1)$
\ENSURE  $\rho^{i}_{th,n}(t)$, $\alpha^{i}(t)$, $\beta^{i}(t)$, $\theta^{i}(t)$
\IF {$\rho^{i}_{r}(t) \neq NaN$}
\STATE $\rho^{i}_{EW\!MA,r}(t)=\omega\cdot\rho^{i}_{r}(t)+(1-\omega)\cdot\rho^{i}_{EW\!MA,r}(t-1)$
\ELSE
\STATE $\rho^{i}_{EW\!MA,r}(t)=\omega\cdot(\rho^{i}_{EW\!MA,n}(t)+\gamma^{min}_{SF})+(1-\omega)\cdot\rho^{i}_{EW\!MA,r}(t-1)$
\ENDIF
\STATE$\rho^{i}_{th,n}(t) = \rho^{i}_{EW\!MA,r}(t) - \gamma^{min}_{SF}$
\IF {($\rho^{i}_{n}(t) \le \rho^{i}_{th,n}(t)$)}
\STATE $\alpha^{i}(t)$ = $\Omega\cdot\alpha^{i}(t-1)$ + 1 
\ELSE
\STATE $\beta^{i}(t)$ = $\Omega\cdot\beta^{i}(t-1)$ + 1 
\ENDIF
\STATE $\theta^{i}(t)$ = $Beta$($\alpha^{i}(t)$, $\beta^{i}(t)$)
\RETURN $\rho^{i}_{th,n}(t)$
\end{algorithmic} 
\end{algorithm}

\subsubsection{Sensitivity to Initial Noise Threshold}

%{\color{red} This subsection is for Noise Threshold, not for Channel Exploration. Do you think we should move it to the end of 3.1?}

Since our solution cannot assume knowledge of channel quality information initially, its initial noise threshold $\rho^{i}_{th,n}(0)$ is unknown. In other words, we do not assume in general a prior estimate of the noise threshold for each channel. Thus, we consider three different schemes for initialization of the noise threshold. 

\noindent\textbf{Pessimistic}: The pessimistic scheme assumes the presence of strong interference for all channels. Consequently, a higher initial noise threshold is assigned to account for this anticipated strong interference across all channels. This approach takes a conservative approach by setting a more cautious initial noise threshold to ensure robustness against potential high levels of interference.  

\noindent\textbf{Optimistic}: The optimistic scheme assumes the absence of interference for all channels. As a result, a low initial noise threshold is assigned, reflecting the expectation of minimal or no interference. This approach takes an optimistic stance by setting a lower initial noise threshold, indicating an assumption of favorable channel conditions with minimal interference.

\noindent\textbf{Seeded}: The seeded scheme utilizes a single sample, typically the first active measurement from a channel, to calculate its initial noise threshold. This initial noise threshold value is then replicated for all other channels, to heuristically serve as their initial noise threshold. This heuristically seeds the process of channel exploration and measurement for the other channels.

\subsection{Channel Exploration}
\emph{Channel Exploration} is the other central component in the design PAMLR, as shown in the top-right block in Figure \ref{fig:illustration}.
PAMLR employs a MAB algorithm specifically designed for efficient channel exploration. Unlike classic MAB algorithms that aim to maximize $\theta^{i}(t)$ during the learning process and strike a balance between exploration and exploitation, our approach focuses on pure exploration in the fixed budget \cite{Pure_Exploration, Best_Arm_Identification}, aiming to find the arm(s) with the maximum expected $\theta^{i}(t)$.

Furthermore, we utilize Thompson sampling \cite{Thompson1933, Thompson1935}, which is capable of selecting a majority of channels since the $\theta^{i}(t)$ are based on the beta distribution. As time progresses, Thompson sampling gradually narrows down the selection. During the initial deployment of nodes, when information about all frequencies is limited, Thompson sampling allows for progressive refinement of channel selection. Hence, Thompson sampling proves to be a suitable method compared to e-greedy and Upper-Confidence Bound (UCB) \cite{UCB} approaches.

Following this procedure, only a few channels (aka ``arms'') are evaluated and ultimately chosen. The channel exploration  algorithm utilizes the noise threshold obtained from the noise threshold update algorithm, along with passive information. We measure only 2$k$ channels for subsequently downselecting the top-$k$ candidate channels. Both the selection of the 2$k$ channels and the top-$k$ channels is based the decreasing order of the $\theta^{i}(t)$.

If the noise level exceeds the noise threshold, we assume that the current noise will have a more detrimental impact on communication with a higher likelihood. To compute $\theta^{i}(t)$, we use $\alpha^{i}(t)$ and $\beta^{i}(t)$. With each noise measurement, $\alpha^{i}(t)$ or $\beta^{i}(t)$ is increased accordingly. We derive $\theta^{i}(t)$ by utilizing the beta distribution with $\alpha^{i}(t)$ and $\beta^{i}(t)$, and accordingly select the top-$k$ channels. 
%based ones with the highest $\theta^{i}(t)$.

\subsubsection{Adapting to Non-Stationary Channels}

To account for non-stationary channels, the $\theta^{i}(t)$ values are adjusted to minimize the impact of outdated information by employing a discount factor $\Omega$ in (0..1]. This is accomplished by multiplying $\Omega$ with the $\alpha^{i}(t-1)$ and $\beta^{i}(t-1)$ values for all frequencies. For stationary channels, the discount factor is simply 1, since it suffices to increment $\alpha^{i}(t-1)$ or $\beta^{i}(t-1)$ as need be.
%Additionally, in static channel conditions, setting $\Omega$ to 1 can be applied  since the values of $\alpha^{i}(t-1)$ and $\beta^{i}(t-1)$ are incremented. 

%{\color{red} Are some previous paragraphs only for stationary channels? If yes, please specify them out like we did in 3.1.1.}

The algorithm for channel exploration in PAMLR is specified in Algorithm \ref{alg:CE}.

\begin{algorithm}[t]
\caption{Algorithm for Channel Exploration}\label{alg:CE}
\begin{algorithmic}[1]
\renewcommand{\algorithmicrequire}{\textbf{Input:}}
\renewcommand{\algorithmicensure}{\textbf{Output:}}
\REQUIRE $\alpha^{i}(t-1)$, $\beta^{i}(t-1)$, $\theta^{i}(t-1)$, $\rho^{i}_{th,n}(t-1)$
\ENSURE  $C^{k}_{alg}$,$\alpha^{i}(t)$, $\beta^{i}(t)$, $\theta^{i}(t)$
  \FOR {$i \in N$} 
  \IF {$i \in argmax_{j \in J: |J| = 2k } \theta^{j}(t-1)$}
  \STATE Play arm (Collect noise sample) $\rho^{i}_{n}(t)$
  \STATE$\rho^{i}_{EW\!MA,r}(t)=\omega\cdot(\rho^{i}_{EW\!MA,n}(t)+\gamma^{min}_{SF})+(1-\omega)\cdot\rho^{i}_{EW\!MA,r}(t-1)$
  \IF {($\rho^{i}_{n}(t) \le \rho^{i}_{th,n}(t-1)$)}
  \STATE $\alpha^{i}(t)$ = $\Omega\cdot\alpha^{i}(t-1)$ + 1 
  \ELSE
  \STATE $\beta^{i}(t)$ = $\Omega\cdot\beta^{i}(t-1)$ + 1 
  \ENDIF
  \ELSE
  \STATE $\alpha^{i}(t)$ = $\Omega\cdot\alpha^{i}(t-1)$ 
  \STATE $\beta^{i}(t)$ = $\Omega\cdot\beta^{i}(t-1)$
  \ENDIF
  \STATE $\theta^{i}(t)$ = $Beta$($\alpha^{i}(t)$, $\beta^{i}(t)$)
  \ENDFOR
 \RETURN $ argmax_{j \in J: |J| = k } \theta^{j}(t)$  
 
\end{algorithmic} 
\end{algorithm}

\section{Evaluation}
In this section, we assess the performance of PAMLR using two distinct methods: simulation-based evaluation and field-experiment based evaluation. For simulation-based evaluation, we synthetically model channel conditions to test PAMLR for a variety of environmental scenarios. For the field-experiment based evaluation, we rely on collecting LoRa network data from environments in different cities to test the efficacy of PAMLR.

%%% AA: Consider saying something more about what is evaluated.  

\subsection{Simulation-based Evaluation}

\subsubsection{Scenarios with Different Channel Environments}
Our simulations incorporate three distinct scenarios that mimic aspects of real-world conditions but also eschew their unknowns, thus allowing us to control our evaluation of specific properties of PAMLR. Specifically, we generate simulation data for specified conditions across all channels, namely, of external interference variability (in Scenario A), of fading variability (in Scenario B), or of composite variability (in Scenario C). These corner cases yield a baseline for comparison and allow us to gain insights into the algorithm's performance under controlled and reproducible conditions.  Given the probabilistic nature of the MAB algorithm with Thompson sampling and its ability to provide varying results in each trial, we conduct each test 500 times. This approach ensured good estimation of $R_{SNR}$ results, enhancing the robustness of our evaluation across different scenarios. 

%%AA: what does enhancing the robustness of our evaluation across different scenarios mean?  is this technically true?

\vspace*{1mm}
\noindent\textbf{Scenario A:} In this scenario, we investigate the performance of our algorithms in the presence of {\em different levels of interference while maintaining similar fading levels across all channels}. We consider a total of 10 channels, out of which 2 channels experience low interference levels, while the remaining 8 channels encounter high interference levels. Our objective is to evaluate PAMLR's ability to handle diverse interference and effectively mitigate their impact.

\begin{figure}
    \centering
    \begin{minipage}[t]{.5\linewidth}
    \centering    \includegraphics[width=0.77\linewidth, angle=-90]{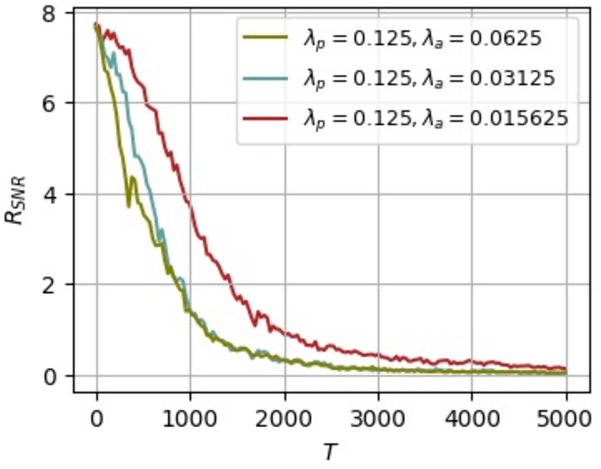}
    \subcaption{Different $\lambda_{a}$}\label{fig:SA_active}
    \end{minipage}%
    \begin{minipage}[t]{.5\linewidth}
    \centering
    \includegraphics[width=0.77\linewidth, angle=-90]{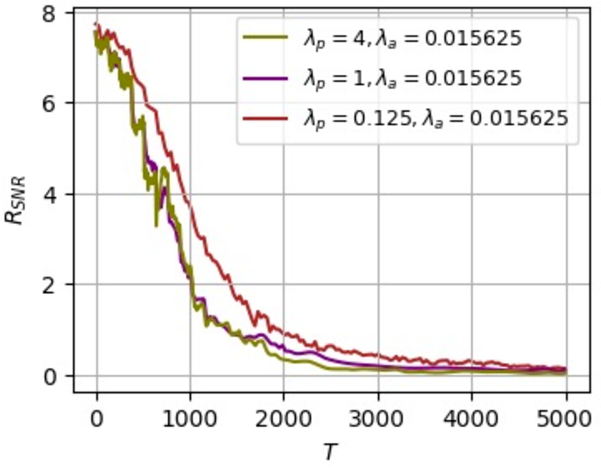}
    \subcaption{Different $\lambda_{p}$}\label{fig:SA_passive}
    \end{minipage}
    
    \caption{The results of $R_{SNR}$ in Scenario A with different $\lambda_{a}$ and $\lambda_{p}$ values, using parameters $k=2$, $N=10$, $\Omega=0.99$, $\omega=0.9$, and a pessimistic initial noise threshold ($\rho^{init}_{th,n} = -80$)}
    \label{fig:SA_active_passive}
\vspace*{-15pt}
\end{figure}

The outcomes of this scenario are depicted in Figure \ref{fig:SA_active_passive}. %In Figure \ref{fig:SA_active}, we observe that increasing the value of $\lambda_{a}$ from 0.015625 to 0.03125 results in an accelerated decrease in $R_{SNR}$. However, further increasing $\lambda_{a}$ from 0.03125 to 0.0625 does not lead to a significant decrease in $R_{SNR}$. Likewise, Figure \ref{fig:SA_passive} illustrates that increasing $\lambda_{p}$ from 0.125 to 0.5 also accelerates the decrease in $R_{SNR}$. However, increasing $\lambda_{p}$ from 0.5 to 2 does not result in a significant decrease in $R_{SNR}$. 
These findings indicate that {\bf the joint selection of $\lambda_{a}$ and $\lambda_{p}$ can converge to low (in this case, near-zero) $R_{SNR}$ values, albeit at varying rates}. A higher $\lambda_{a}$ yields earlier identification of the correct noise thresholds (cf.~Figure \ref{fig:SA_active_passive}(a)), but requires increased energy expenditure. A higher $\lambda_{p}$ also accelerates the discovery of the correct top-$k$ channels (cf.~Figure \ref{fig:SA_active_passive}(b)), with more modest increase in energy consumption.  While higher $\lambda_{a}$ and $\lambda_{p}$ result in quicker convergence, there appears to be a level beyond which further increases in these parameters do not significantly reduce $R_{SNR}$ any further. Consequently, with respect to energy consumption, it is more favorable to select $\lambda_{a} = 0.03125$ instead of $0.0625$ (cf.~ Figure \ref{fig:SA_active_passive}(a)), and $\lambda_{p} = 1$ instead of $4$ (cf.~Figure \ref{fig:SA_active_passive}(b)).

More importantly, {\bf choosing a higher $\lambda_{p}$ can compensate for a smaller $\lambda_{a}$, i.e., achieve similar regret performance while being more energy efficient}: note that the results achieved with $\lambda_{p} = 4$ and $\lambda_{a} = 0.015625$ are similar to those obtained with $\lambda_{p} = 0.125$ and $\lambda_{a} = 0.0625$. %This illustrates that {\bf increasing $\lambda_{p}$ while maintaining a low $\lambda_{a}$ can be energy efficient}, while providing similar performance to the case with low $\lambda_{p}$ and high $\lambda_{a}$.

\vspace*{1mm}
\noindent\textbf{Scenario B:} In this scenario, we investigate a configuration comprising 10 channels characterized by {\em minimal interference but differing fading}. More specifically, 2 channels exhibit negligible fading, while the remaining 8 channels encounter significant fading. The intent is to analyze the influence of frequency-selective fading conditions on the effectiveness of PAMLR, while ensuring consistent levels of interference across all channels.

\begin{figure}
    \centering
    \begin{minipage}[t]{.5\linewidth}
    \centering
    \includegraphics[width=0.76\linewidth, angle=-90]{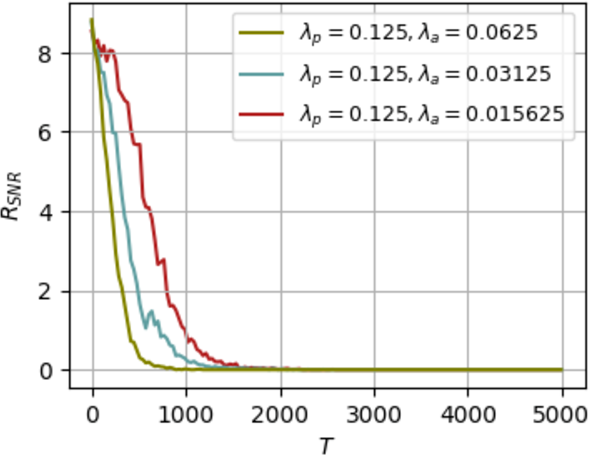}
    \subcaption{Different $\lambda_{a}$}\label{fig:SB_active}
    \end{minipage}%
    \begin{minipage}[t]{.5\linewidth}
    \centering
    \includegraphics[width=0.76\linewidth, angle=-90]{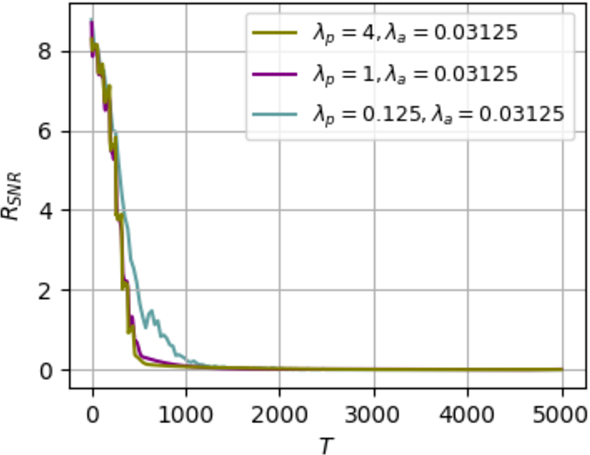}
    \subcaption{Different $\lambda_{p}$}\label{fig:SB_passive}
    \end{minipage}
    \caption{The results of $R_{SNR}$ in Scenario B with different $\lambda_{a}$ and $\lambda_{p}$ values, using parameters $k=2$, $N=10$, $\Omega=0.99$, $\omega=0.9$, and a pessimistic initial noise threshold ($\rho^{init}_{th,n} = -80$)}
    \label{fig:SB_active_passive}
    \vspace{-5pt}
\end{figure}

The outcomes of Scenario B are depicted in Figure \ref{fig:SB_active_passive}.
The findings are similar to those of Scenario A, albeit the increase of $\lambda_{a}$ has a more pronounced rate of decrease in $R_{SNR}$ than the increase of $\lambda_{p}$.  This is partly because Scenario B converges more quickly than Scenario A to low $R_{SNR}$.
It turns out that this is due to the assumption of a pessimistic initial noise threshold, which for Scenario A implies that the noise threshold for a channel typically yields a high estimation error until active sampling has occurred for the channel; in contrast, in Scenario B, active sampling is necessary only for those channels with low fading, and thus it converges faster.

\vspace*{1mm}
\noindent\textbf{Scenario C:} This scenario investigates a {\em combination of diverse interference and fading conditions across all 10 channels}. By subjecting PAMLR to these combined conditions, the algorithm's resilience in extreme cases is demonstrated, which yields confidence for environments typically encountered in the field. 

\begin{figure}
    \centering    
    \begin{minipage}[t]{.5\linewidth}
    \centering
    \includegraphics[width=0.77\linewidth, angle=-90]{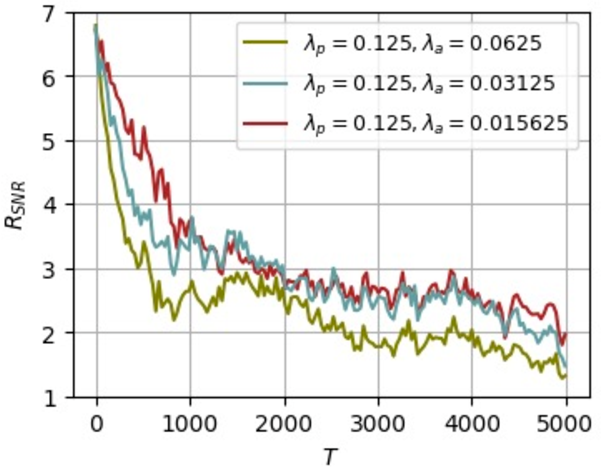}
    \subcaption{Different $\lambda_{a}$}\label{fig:SC_active}
    \end{minipage}%
    \begin{minipage}[t]{.5\linewidth}
    \centering
    \includegraphics[width=0.77\linewidth, angle=-90]{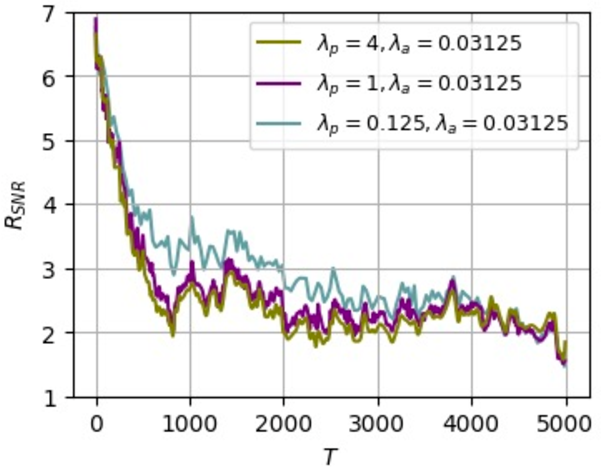}
    \subcaption{Different $\lambda_{p}$}\label{fig:SC_passive}    \end{minipage}    
    \caption{The results of $R_{SNR}$ in Scenario C with different $\lambda_{a}$ and $\lambda_{p}$ values, using parameters $k=2$, $N=10$, $\Omega=0.99$, $\omega=0.9$, and a pessimistic initial noise threshold ($\rho^{init}_{th,n} = -80$)}
    \label{fig:SC_active_passive}
\end{figure}

The outcomes of Scenario C are depicted in Figure \ref{fig:SC_active_passive}. As before, Figure \ref{fig:SC_active_passive}(a) demonstrates that increasing the value of $\lambda_{a}$ leads to a more pronounced rate of decrease in $R_{SNR}$. 
%Like Scenario B, enhancing the value of $\lambda_{p}$ does not yield significant improvements in $R_{SNR}$. 
In contrast to the other scenarios, until $T$ = 5000 Scenario C does not converge to a near zero $R_{SNR}$. This highlights the impact of composite interference and fading characteristics on the performance: finding the proper noise threshold and rewards for the composite scenario requires conducting more active and passive measurements than in Scenarios A and B, {\bf leading to a longer convergence time requirement} with similar parameters as before.

\subsubsection{Convergence relative to Initial Noise Threshold}
To examine the impact of different initial noise thresholds, we again consider Scenario C given its richer variability.  To contrast the use of an initial noise threshold of -80dBm for all channels (``pessimistic''), we also simulate for cases where the initial noise threshold is -120dBm for all channels (``optimistic'') and where it is identically initialized for all channels based on the first active measurement of some randomly chosen channel (``seeded''). 
%present a scenario where all channels encounter diverse interference and fading conditions, while maintaining comparable levels. This implies that the initial noise threshold $\rho_{th,n}^i(0)$ for the first active measurement in the seeded scheme closely aligns with $\rho_{th,n}^i(0)$ across all channels. For the pessimistic and optimistic schemes, we assign initial noise thresholds of -80 and -120, respectively. 

\begin{figure}[t]
    \centering	\includegraphics[width=0.44\linewidth, trim = 9 0 0 0, angle=-90]{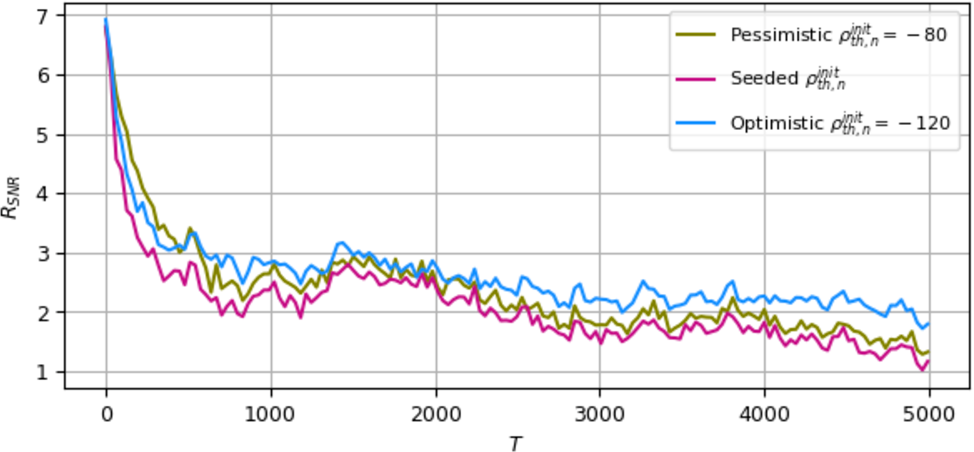}
    \caption{Comparing the results of $R_{SNR}$ in Scenario C for 3 different initial noise thresholds: pessimistic ($\rho^{init}_{th,n} = -80$) vs. seeded vs. optimistic ($\rho^{init}_{th,n} = -120$). The other parameters are $k=2$, $N=10$, $\Omega=0.99$, $\omega=0.9$, $\lambda_a=0.0625$, and $\lambda_p=0.125$}
    \label{fig:diff_init_noise_th}	
\end{figure}

Figure \ref{fig:diff_init_noise_th} depicts the performance of these three initial noise threshold selection schemes. Although the optimistic scheme initially has superior $R_{SNR}$ compared to the pessimistic scheme, it converges more slowly. 
%As time progresses, the optimistic scheme exhibits a gradual decrease in regret, indicating an improvement in performance. In contrast, the pessimistic scheme shows a steady decrease in regret. 
%Eventually, as the $R_{SNR}$ stabilizes, the regret of the pessimistic scheme becomes lower than that of the optimistic scheme. This suggests that once the system reaches a stable state, the pessimistic scheme achieves better performance in terms of reducing regret.
Notably, {\bf the seeded scheme demonstrates the best results in terms of convergence and regret performance}. This finding holds for other simulations and data sets we considered.  The heuristic of baselining the Noise Threshold with even a limited amount of information about some channel, as opposed to being a priori pessimistic or optimistic about all channels appears to be the most promising option for application.

%These findings highlight the importance of the initial noise threshold on the performance of methods. By examining different schemes and their corresponding regret trends, we can gain insights into the system's behavior and make informed decisions regarding the selection of an appropriate initial noise threshold.

\subsection{Field Evaluation}
\begin{figure} [t]
    \centering
	\subcaptionbox{\label{fig3:a}}{
        \includegraphics[width=1.16in, angle=90]{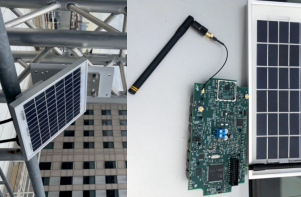}}
        \subcaptionbox{\label{fig3:b}}{
	\includegraphics[width=1.16in, angle=90]{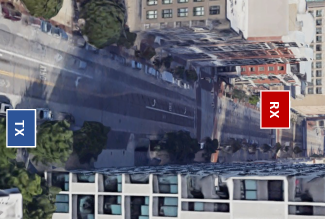}}
 	\subcaptionbox{\label{fig3:c}}{
        \includegraphics[width=1.16in, angle=90]{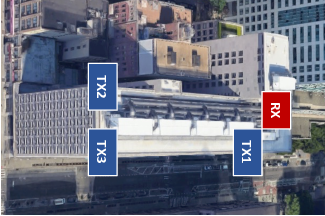}}
  	\subcaptionbox{\label{fig3:d}}{
	\includegraphics[width=1.16in, angle=90]{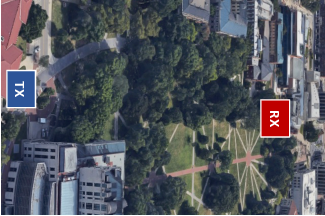}}
~		 \vspace{-2mm}
	\caption{(a)  MKII devices used for field testing; upper image shows unboxed unit, lower image shows boxed unit under solar panel. The locations of the field tests conducted in (b) Downtown Columbus, (c) Downtown Brooklyn, and (d) Oval Park}
	\label{fig:urban_settings}
	\vspace{-2mm}
\end{figure}

For more realistic evaluation, we collect data in three distinct  urban settings: Downtown Columbus located in the Midwest region of the United States  (Figure \ref{fig:urban_settings}(b)), Downtown Brooklyn located in the Northeastern region of the United States, (Figure \ref{fig:urban_settings}(c)), and Oval Park, an urban city park (Figure \ref{fig:urban_settings}(d)). These settings represent urban environments with a diverse range of interference and fading characteristics. Rather than execute PAMLR in real-time, we leverage the collected data to assess its effectiveness retrospectively. 

To gather field data, we employ MKII (``Mach 2'') devices \cite{MKII} equipped with a Semtech SX1276 LoRa module, depicted in Figure \ref{fig:urban_settings}(a). For our data collection, we configure the LoRa settings with a SF of 8, bandwidth of 500 KHz, and coding rate of 4/5. The frequency range for our data collection is the ISM band between 902 MHz and 928 MHz. 
%{\color{blue} In the field evaluation, $T$ represents the total number of samples for each frequency throughout the entire duration. However, it's important to note that the duration of the field evaluation may vary.}
As with the simulation-based evaluation, we conduct each field evaluation test 500 times to ensure good, robust estimation of $R_{SNR}$.

\vspace*{1mm}
\noindent\textbf{Downtown Columbus}:  In this setting with modestly tall buildings, we deploy one transmitter and one receiver on ground-level poles. The total duration of this evaluation spans 30 minutes. There is substantial presence of road traffic, which presents time-varying obstacles and reflectors between the transmitter and receiver. The setting thus lets us {\em explore the impact of dynamic fading} among other environmental factors. 
%The results obtained in this location are influenced by these factors, including the effects of vehicle movement on signal propagation and the resulting variations in fading characteristics.

%\begin{figure}[b]%
%	\centering
%	\includegraphics[width=0.34\textwidth, trim = 9 0 0 0, angle=90]{./images/cond_columbus_downtown.eps}
%	\caption{Channel Condition Columbus Downtown}
%	\label{network_snr}	
%\end{figure}

\begin{figure}
    \centering	             
    \includegraphics[width=0.70\linewidth, trim = 12 0 0 0, angle=-90]{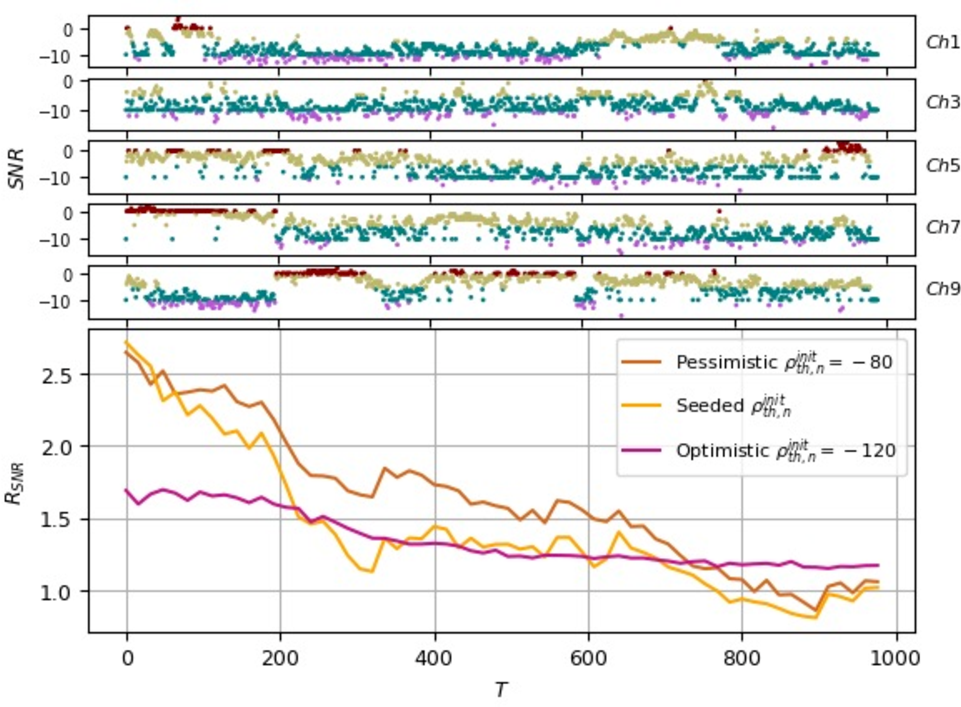}
    \caption{The channel condition and the results of $R_{SNR}$ for different $\rho^{init}_{th,n}$ values, using parameters $k=2$, $N=10$, $\Omega=0.99$, $\omega=0.9$, $\lambda_{a}=0.03125$, and $\lambda_{p}=0.25$, were obtained from measurements conducted in Downtown Columbus setting}
    \label{fig:CD_cond}
\end{figure}

Figure \ref{fig:CD_cond} represents the channel conditions in Downtown Columbus, wherein we depict only 5 channels given space limitations but which suffice to show channel conditions vary significantly over time (with the exception of Ch3). The figure also shows the evolution of $R_{SNR}$ when using different $\rho_{th,n}^{init}$ values. %In the figure depicting the channel conditions, we observe that, except for channel 3, most of the channels exhibit varying channel conditions over time.

When analyzing the convergence of $R_{SNR}$ with different $\rho_{th,n}^{init}$ values, just as in the case of Scenario C, we find that the optimistic $\rho_{th,n}^{init}$ initially provides a lower $R_{SNR}$ value but over time converges slower than the pessimistic and seeded $\rho_{th,n}^{init}$ values. Eventually (i.e., $T=700$), the pessimistic and seeded $\rho_{th,n}^{init}$ values yield lower $R_{SNR}$ values.

\begin{figure}
    \centering
    \begin{minipage}[t]{.5\linewidth}
    \centering
    \includegraphics[width=0.75\linewidth, angle=-90]{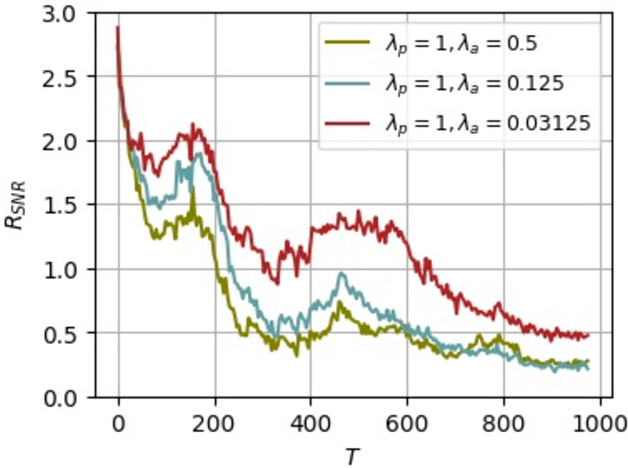}
    \subcaption{Different $\lambda_{a}$}\label{subfig:CD_active}
    \end{minipage}%
    \begin{minipage}[t]{.5\linewidth}
    \centering
    \includegraphics[width=0.75\linewidth, angle=-90]{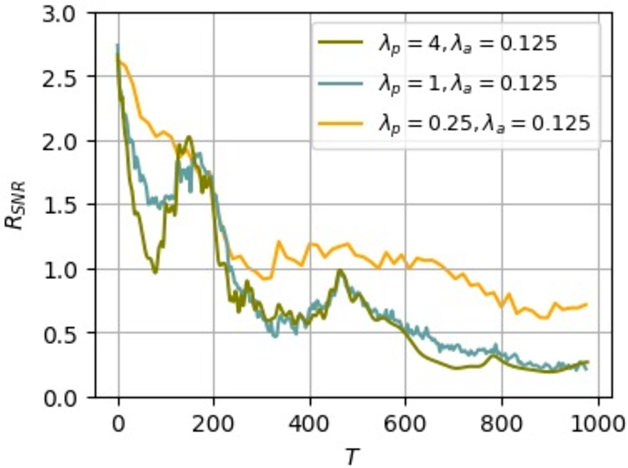}
    \subcaption{Different $\lambda_{p}$}\label{subfig:CD_passive}
    \end{minipage}
    
    \caption{The results of $R_{SNR}$ for  different $\lambda_{a}$ and $\lambda_{p}$ values, using parameters $k=2$, $N=10$, $\Omega=0.99$, $\omega=0.9$, and a seeded $\rho^{init}_{th,n}$, were obtained from measurements conducted in Downtown Columbus setting.}
    \label{fig:CD_active_passive}
\end{figure}

Figure~\ref{fig:CD_active_passive} shows the convergence of $R_{SNR}$ with different values of $\lambda_{a}$ and $\lambda_{p}$, again corroborating the main findings of the simulated scenarios. Increasing both $\lambda_{a}$ and $\lambda_{p}$ leads to lower $R_{SNR}$ (cf.~Figure~\ref{fig:CD_active_passive}(a)); also, it is possible to compensate for lowered $\lambda_{a}$ with increased $\lambda_{p}$ (cf.~the green curves in ~Figure~\ref{fig:CD_active_passive}). %REWRITE COMPENSATION

%With $\lambda_{p} = 1$ and $\lambda_{a} = 0.5$, it takes $T=240$ to reach an $R_{SNR}$ value of 0.5. On the other hand, when $\lambda_{p} = 4$ and $\lambda_{a} = 0.03125$, it takes $T=700$ to achieve the same $R_{SNR}$ value. This indicates that increasing $\lambda_{p}$ while keeping $\lambda_{a}$ low can bring the $R_{SNR}$ close to the levels achieved with high $\lambda_{p}$, but it requires more time to reach the desired $R_{SNR}$ level.

\vspace*{1mm}
\noindent\textbf{Downtown Brooklyn}: In Downtown Brooklyn setting with tall urban canyons, we deploy one transmitter and three receivers on the rooftops of buildings. In contrast to Downtown Columbus, this setting exhibits {\em a mix of channels with both low and high SNR and most of the channels demonstrate varying SNR levels over time}. 

Unlike the other field evaluations, we were able to collect week-long data in this setting, but due to platform constraints, the data is not raw but rather statistical: it was not possible to store the data directly at the receiver side, so the receivers were instrumented to transmit the statistical information back to the transmitter over the low capacity LoRa network. The statistical data includes metrics such as the 95th percentile of noise, 5th percentile of SNR, and 5th percentile of RSSI, which  individually are derived from 100 raw data samples. This limitation let us {\em evaluate the effectiveness of PAMLR --even if with greater $R_{SNR}$ levels--- in the context of statistical aggregation of samples.} The total duration of this evaluation spanned 92 hours.

\begin{figure}[t]
	\centering	\includegraphics[width=0.71\linewidth, trim = 12 0 0 0, angle=-90]{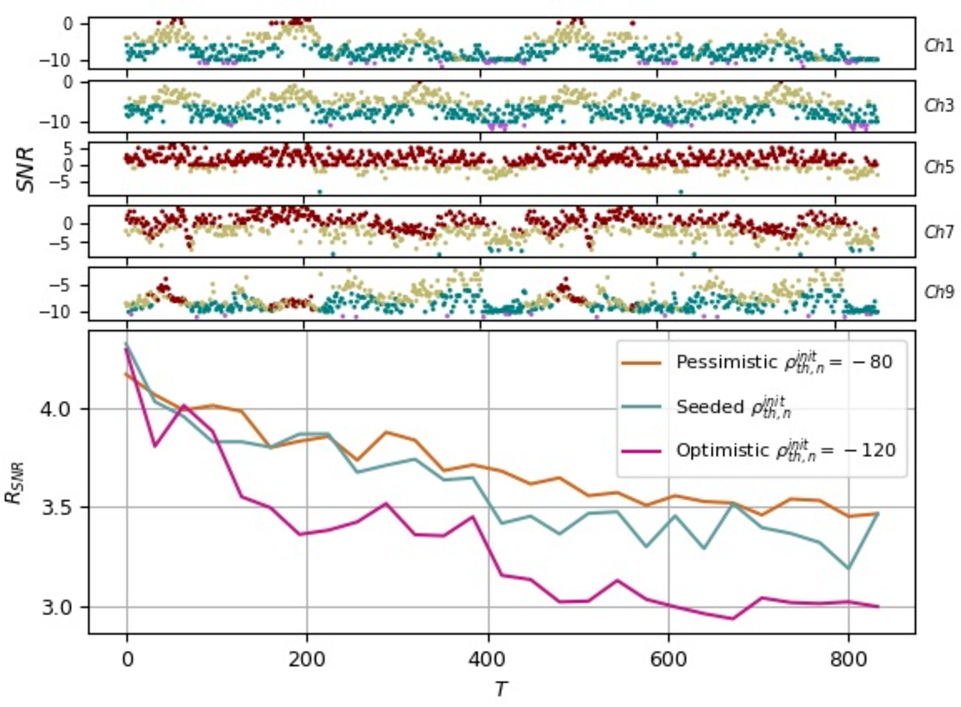}
	\caption{The channel condition and the results of $R_{SNR}$ for different $\rho^{init}_{th,n}$ values, using parameters $k=2$, $N=10$, $\Omega=0.99$, $\omega=0.9$, $\lambda_{a}=0.03125$, $\lambda_{p}=0.125$, were obtained from RX3 located in Downtown Brooklyn setting}
	\label{fig:NYC_cond}
\end{figure}

Figure \ref{fig:NYC_cond} shows convergence of $R_{SNR}$ with different $\rho_{th,n}^{init}$ values that, because of the statistical aggregation, is slower than in the other scenarios. The optimistic $\rho_{th,n}^{init}$ again provides lower $R_{SNR}$ values to begin with, but because of the aggregation, does not crossover with the pessimistic and seeded initializations within the experiment, although we expect it would given more time.

%and end of the sample points. It is important to recall that these results are obtained from normalization data, and the sample points used are relatively few. Based on this, with more sample points, it is expected that the pessimistic and seeded $\rho_{th,n}^{init}$ values would yield even lower $R_{SNR}$ values compared to the optimistic case. The larger sample size would likely provide a more comprehensive representation of the data and further highlight the differences in $R_{SNR}$ between different $\rho_{th,n}^{init}$ values.

\begin{figure}[t]
    \centering
    \begin{minipage}[t]{.5\linewidth}
    \centering    \includegraphics[width=0.74\linewidth, angle=-90]{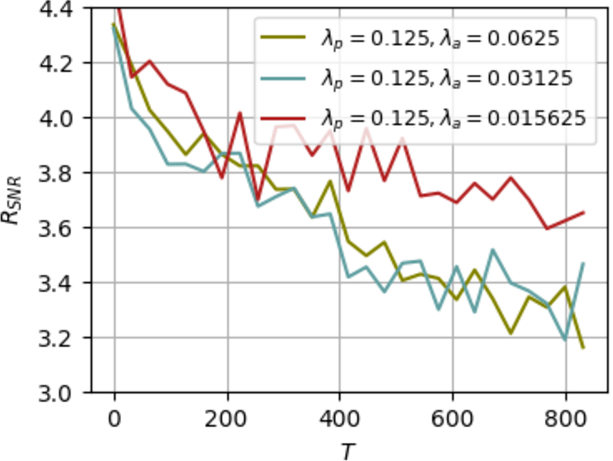}
    
    \subcaption{Different $\lambda_{a}$}\label{subfig:NYC_active}
    \end{minipage}%
    \begin{minipage}[t]{.5\linewidth}
    \centering
    \includegraphics[width=0.74\linewidth, angle=-90]{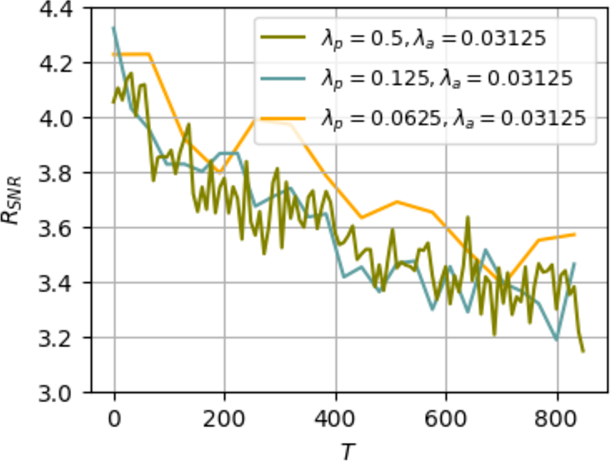}
    \subcaption{Different $\lambda_{p}$}\label{subfig:NYC_passive}
    \end{minipage}
    
    \caption{The results of $R_{SNR}$ for  different $\lambda_{a}$ and $\lambda_{p}$ values, using parameters $k=2$, $N=10$, $\Omega=0.99$, $\omega=0.9$, and a seeded $\rho^{init}_{th,n}$, were obtained from RX3 located in Downtown Brooklyn setting}
    \label{fig:NYC_active_passive}
\end{figure}

Figure \ref{fig:NYC_active_passive} shows convergence of   $R_{SNR}$ for different $\lambda_{a}$ and $\lambda_{p}$. The broad trends are similar to other scenarios: increasing both leads to lower $R_{SNR}$ values, and the green curves show how to compensate for a lower $\lambda_{a}$ value with a higher $\lambda_{p}$ value. Nevertheless, the slow convergence implies that the $R_{SNR}$ values do not converge to the low levels of the other scenarios given the number of statistical aggregate samples in this experiment.

%. However, it is worth noting that the $R_{SNR}$ values in this city are larger compared to other cities. This suggests that the algorithm is able to reduce $R_{SNR}$ using the normalization data, but the reduction is not as significant as with raw data.

\vspace*{1mm}
\noindent\textbf{Oval Park}:  For this mixed natural-cum-builtup setting, we focus on evaluating {\em the impact of different values of $k$ and $N$}. As we did in the field evaluation in the Downtown Columbus setting, we deploy one transmitter and one receiver on ground-level poles. The entire evaluation lasts for a duration of 30 minutes. The channel conditions in this setting are characterized by two channels having low SNR, while the other channels exhibit high SNR. 

\begin{figure}[t]
    \centering
    \begin{minipage}[t]{.5\linewidth}
    \centering
    \includegraphics[width=0.73\linewidth, angle=-90]{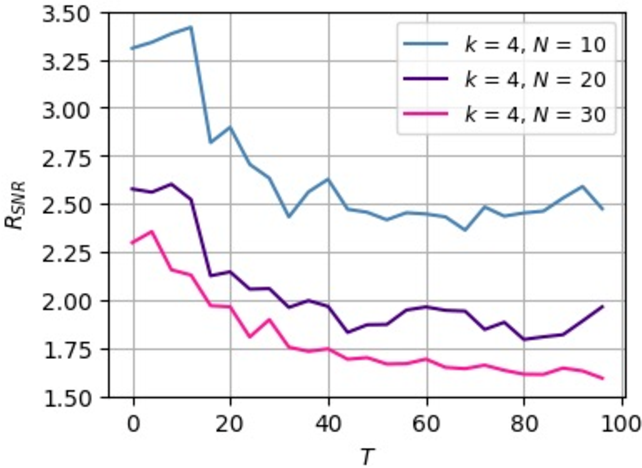}
    \subcaption{Different $N$}\label{subfig:oval_N}
    \end{minipage}%
    \begin{minipage}[t]{.5\linewidth}
    \centering
    \includegraphics[width=0.73\linewidth, angle=-90]{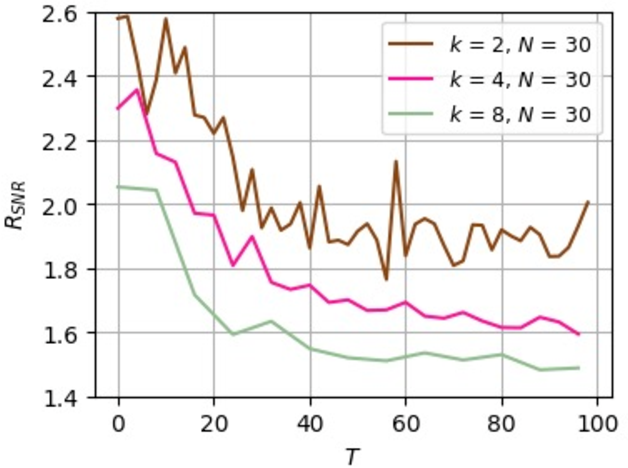}
    \subcaption{Different $k$}\label{subfig:oval_k}
    \end{minipage}
    
    \caption{The results of $R_{SNR}$ for different $N$ and $k$ values, using parameters $\Omega=0.99$, $\omega=0.9$, $\lambda_{a}=1$, $\lambda_{p}=2$, and a seeded $\rho^{init}_{th,n}$, were obtained in the Oval Park Setting}
    \label{fig:oval_k_N}
\end{figure}

Figure \ref{fig:oval_k_N} illustrates the convergence of $R_{SNR}$ for different combinations of $N$ and $k$ values. We observe that increasing $N$ leads to lower $R_{SNR}$ values. This can be attributed to the channel conditions in the park setting, where having a smaller $N$ increases the possibility of selecting two channels with low SNR, resulting in poorer overall $R_{SNR}$ performance. Interestingly, Figure~\ref{fig:oval_k_N} demonstrates that increasing $k$ leads to quicker convergence to lower $R_{SNR}$ values: this is because it takes longer to determine the top-$k$ channels for smaller $k$.

\section{PAMLR with Dynamic Sampling Rate}

\noindent\textbf{Adaptive selection of $\lambda_{a}$ and $\lambda_{p}$}: 
We discuss an adaptive mechanism for PAMLR to dynamically adjust the values of $\lambda_{a}$ and $\lambda_{p}$ in an online fashion in response to the current channel condition. Adaptation is desirable on one hand since diverse channel conditions may not require high values of $\lambda_{a}$ and $\lambda_{p}$ and higher than need be values yield increased energy consumption. On the other hand, in the presence of channel dynamics, increasing the values of $\lambda_{a}$ and $\lambda_{p}$ may be warranted to accurately select the top-$k$ channels for the changed channel conditions.

%We therefore consider the incorporation of an adaptive mechanism to dynamically fine-tune the values of $\lambda_{a}$ and $\lambda_{p}$ based on observed channel conditions. 

The mechanism design uses frequency-specific estimation of the variation in the passive and active channel measurements. A minimal variance between the previous and current measurement results indicates channel stability and, in such a situation, reducing the values of $\lambda_{p}$ and $\lambda_{a}$ can lead to energy reduction without loss of accuracy. Conversely, a substantial variance raises suspicion of a change in the set of top-$k$ channels and, in such a situation, increasing the values of $\lambda_{p}$ and $\lambda_{a}$ enables responsive discovery of the true top-$k$ channels.

More specifically, we program the adaptivity of $\lambda_{a}$ based on the variation of the SNR obtained through active measurements:  When adjusting $\lambda_{a}$ for each active measurement at time $t$, we calculate the aggregated EWMA SNR value from the top-$k$ channels, denoted as $\gamma^{Top_k}_{EW\!MA}(t)$. This calculation is performed using the following equation:
\vspace*{-0.5mm}
\begin{equation}
\gamma^{Top\_k}_{EW\!MA}(t)= \hat{\omega} \cdot \frac{\sum_{i=1}^k\gamma^{i}(t)}{k} + (1-\hat{\omega}) \cdot \gamma^{Top\_k}_{EW\!MA}(t-1),
\vspace*{1.5mm} \\
\end{equation}
where $\hat{\omega}$ represents the weighting factor in the EWMA calculation for SNR. Analyzing the current $\gamma^{Top\_k}_{EW\!MA}(t)$ in comparison to the previous $\gamma^{Top\_k}_{EW\!MA}(t-1)$ can provide insight into whether the current value of $\lambda_{a}$ should be adjusted upward or downward. For instance, consider situations where the current value is greater than, smaller than, or equal to the previous value. In the first two scenarios, increasing $\lambda_{a}$ can aid in the identification of the top-k channels, while in the latter case, decreasing $\lambda_{a}$ can result in reduced energy consumption.

For adapting $\lambda_{p}$, during each channel exploration, the $\alpha^{Top\_2k}_{EW\!MA}(t)$ and $\beta^{Top\_2k}_{EW\!MA}(t)$ is calculated, where $\alpha^{Top\_2k}_{EW\!MA}$ and $\beta^{Top\_2k}_{EW\!MA}$ respectively represent the EWMA of the total number of noise values below and above (or equal to) the noise threshold, among the top-$2k$ channels selected during the channel exploration. These values are calculated as follows:
\vspace*{-0.5mm}
\begin{equation}
\alpha^{Top\_2k}_{EW\!MA}(t)= \check{\omega}\cdot S(t) + (1-\check{\omega}) \cdot \alpha^{Top\_k}_{EW\!MA}(t-1)
\end{equation}
\vspace*{1mm}
\begin{equation}
\beta^{Top\_2k}_{EW\!MA}(t)= \check{\omega}\cdot(2k-S(t)) + (1-\check{\omega}) \cdot \beta^{Top\_k}_{EW\!MA}(t-1) 
\vspace*{1.5mm} \\
\end{equation}
Here, $S(t)$ represents the total number of noise values below the noise threshold in the current result of channel exploration from the $2k$ channels, and $\check{\omega}$ represents the weighting factor in the EWMA calculation for $\alpha^{Top\_2k}_{EW\!MA}(t)$ and $\beta^{Top\_2k}_{EW\!MA}(t)$.  From these two, the value of $\theta^{Top\_2k}_{EW\!MA}(t)$ is calculated using:
\vspace*{-.5mm}
\begin{equation}
\theta^{Top\_2k}_{EW\!MA}(t)= \frac{\alpha^{Top\_2k}_{EW\!MA}(t)}{\alpha^{Top\_2k}_{EW\!MA}(t)+\beta^{Top\_2k}_{EW\!MA}(t)}
\end{equation}
%Similar to $\gamma^{Top\_k}_{EW\!MA}(t)$, 
In this case, the variance between $\theta^{Top\_2k}_{EW\!MA}(t)$ and the current $\theta^{Top\_}$ $^{2k}(t)$ obtained from all $2k$ channels selected during each channel exploration is used to adapt $\lambda_{p}$. (Note that unlike $\theta^{i}(t)$ in  channel exploration, which employs the beta distribution, $\theta^{Top\_2k}_{EW\!MA}(t)$ needs to be calculated using the rate $\frac{\alpha^{i}(t)}{\alpha^{i}(t)+\beta^{i}(t)}$ without relying on the beta distribution; this is necessary because the beta distribution introduces randomness, making it unsuitable for stable comparisons between current and previous values). Similar to the adapting $\lambda_{a}$, comparing the present $\theta^{Top\_2k}_{EW\!MA}(t)$ to the prior $\theta^{Top\_2k}_{EW\!MA}(t-1)$ can assist in determining whether the current value of $\lambda_{p}$ should be adjusted upwards or downwards. For example, consider situations where the current value is greater than, smaller than, or equal to the previous value. In the first two scenarios, increasing $\lambda_{p}$ can improve the identification of the top-k channels, while in the latter case, decreasing $\lambda_{p}$ can lead to reduced energy consumption.

\section{Conclusion}

We have demonstrated that PAMLR achieves energy efficient selection of the top-most reliable channels in urban environments, notwithstanding their complex dynamics of both external interference and fading. Active measurements on a channel are used primarily to estimate a fading- and interference- aware noise threshold of the current set of top-most reliable channels, and the noise threshold is used along with interference-aware passive measurements on the channel in the ranking of its reliability.

%To efficiently achieve low duty cycle operation in low-power wireless networks within urban environments, which are susceptible to the complex and variable dynamics of external interference and fading, PAMLR is specifically designed to achieve energy-efficient channel selection. 

%This is accomplished through a meticulous balance of active and passive channel measurements. Active channel measurements are finely adjusted to a low level to update noise thresholds, while passive channel measurements are set to an appropriate high level to select the most promising channels using noise threshold-guided channel exploration. 

The efficiency also derives from a channel exploration strategy for passive sampling that is based on MAB and, more specifically, the MAB style of pure exploration with a fixed budget.  Rather than use the more conventional e-greedy or UCB-based MAB approaches, PAMLR leverages Thompson sampling with Beta Distribution; it thus initially provides all frequencies with an equal albeit random opportunity to be selected. 
%It utilizes Thompson sampling with Beta Distribution randomness when the number of samples is small. 
As the exploration progresses, it reduces the number of channel measurements, focusing on exploring only a few ---instead of all--- frequencies.

Extensive evaluations, involving both simulations and field experiments, consistently demonstrate the communication quality achieved by PAMLR, in terms of the SNR regret with respect to an optimal channel allocation policy. 

%Utilizing both passive and active channel measurements results in lower SNR regret, and increasing passive channel measurements while maintaining low active channel measurements can effectively compensate for the high energy costs associated with active measurements. Moreover, PAMLR could exhibit diminished SNR regret even when provided with statistical inputs instead of raw data.

PAMLR as designed can be incorporated into a diverse set of medium access controller (MAC) protocols. In fact, the original motivation for solving this problem came from extending a fielded wireless LoRa mesh network application in a major metropolis from its ultra-low duty cycle but single channel MAC realization to a comparably low duty cycle incarnation that uses multiple channel. The use of reinforcement learning in the fashion presented here was key to achieving our goal. In our related work, reinforcement learning has also yielded efficient local and regional routing policies across graphs with highly variable densities and sizes. In the future, it would be of interest to look at reinforcement learning of energy-efficient cross-layer link and routing layer designs.

\bibliographystyle{ACM-Reference-Format}
\bibliography{main}


\begin{thebibliography}{27}

%%% the Web address.

\ifx \showCODEN    \undefined \def \showCODEN     #1{\unskip}     \fi
\ifx \showDOI      \undefined \def \showDOI       #1{#1}\fi
\ifx \showISBNx    \undefined \def \showISBNx     #1{\unskip}     \fi
\ifx \showISBNxiii \undefined \def \showISBNxiii  #1{\unskip}     \fi
\ifx \showISSN     \undefined \def \showISSN      #1{\unskip}     \fi
\ifx \showLCCN     \undefined \def \showLCCN      #1{\unskip}     \fi
\ifx \shownote     \undefined \def \shownote      #1{#1}          \fi
\ifx \showarticletitle \undefined \def \showarticletitle #1{#1}   \fi
\ifx \showURL      \undefined \def \showURL       {\relax}        \fi
% invisible to TeX
\providecommand\bibfield[2]{#2}
\providecommand\bibinfo[2]{#2}
\providecommand\natexlab[1]{#1}
\providecommand\showeprint[2][]{arXiv:#2}

\bibitem[{Abdelghany} et~al\mbox{.}(2022)]%
        {active_ch_7}
\bibfield{author}{\bibinfo{person}{A. {Abdelghany}}, \bibinfo{person}{B.
  {Uguen}}, \bibinfo{person}{C. {Moy}}, {and} \bibinfo{person}{D. {Lemur}}.}
  \bibinfo{year}{2022}\natexlab{}.
\newblock \showarticletitle{Decentralized Adaptive Spectrum Learning in
  Wireless IoT Networks Based on Channel Quality Information}.
\newblock \bibinfo{journal}{\emph{IEEE Internet of Things Journal}}
  \bibinfo{volume}{9}, \bibinfo{number}{20} (\bibinfo{year}{2022}),
  \bibinfo{pages}{19660--19669}.
\newblock
\urldef\tempurl%
\url{https://doi.org/10.1109/JIOT.2022.3167016}
\showDOI{\tempurl}


\bibitem[Auer et~al\mbox{.}(2002)]%
        {UCB}
\bibfield{author}{\bibinfo{person}{P. Auer}, \bibinfo{person}{N. Cesa-Bianchi},
  {and} \bibinfo{person}{P. Fischer}.} \bibinfo{year}{2002}\natexlab{}.
\newblock \showarticletitle{Finite-time Analysis of the Multiarmed Bandit
  Problem}.
\newblock \bibinfo{journal}{\emph{Machine Learning}}  \bibinfo{volume}{47}
  (\bibinfo{date}{05} \bibinfo{year}{2002}), \bibinfo{pages}{235--256}.
\newblock
\urldef\tempurl%
\url{https://doi.org/10.1023/A:1013689704352}
\showDOI{\tempurl}


\bibitem[Bubeck et~al\mbox{.}(2011)]%
        {Pure_Exploration}
\bibfield{author}{\bibinfo{person}{S. Bubeck}, \bibinfo{person}{R. Munos},
  {and} \bibinfo{person}{G. Stoltz}.} \bibinfo{year}{2011}\natexlab{}.
\newblock \showarticletitle{Pure Exploration in Finitely-armed and
  Continuous-armed Bandits}.
\newblock \bibinfo{journal}{\emph{Theoretical Computer Science}}
  \bibinfo{volume}{412} (\bibinfo{year}{2011}), \bibinfo{pages}{1832--1852}.
\newblock
\urldef\tempurl%
\url{https://doi.org/10.1016/j.tcs.2010.12.059}
\showDOI{\tempurl}


\bibitem[Chiti et~al\mbox{.}(2015)]%
        {passive_ch_4}
\bibfield{author}{\bibinfo{person}{F. Chiti}, \bibinfo{person}{R. Fantacci},
  {and} \bibinfo{person}{A. Tani}.} \bibinfo{year}{2015}\natexlab{}.
\newblock \showarticletitle{Performance Evaluation of an Adaptive Channel
  Allocation Technique for Cognitive Wireless Sensor Networks}. In
  \bibinfo{booktitle}{\emph{2015 IEEE Globecom Workshops (GC Wkshps '15)}}.
  \bibinfo{pages}{1--6}.
\newblock
\urldef\tempurl%
\url{https://doi.org/10.1109/GLOCOMW.2015.7414076}
\showDOI{\tempurl}


\bibitem[{DASH7 Alliance}({[n.\,d.]})]%
        {LPWAN_DASH7}
\bibfield{author}{\bibinfo{person}{{DASH7 Alliance}}.}
  \bibinfo{year}{[n.\,d.]}\natexlab{}.
\newblock \bibinfo{booktitle}{\emph{DASH7 Alliance Protocol}}.
\newblock
\urldef\tempurl%
\url{https://www.dash7-alliance.org/}
\showURL{%
\tempurl}


\bibitem[{Du} and {Roussos}(2012)]%
        {passive_ch_3}
\bibfield{author}{\bibinfo{person}{P. {Du}} {and} \bibinfo{person}{G.
  {Roussos}}.} \bibinfo{year}{2012}\natexlab{}.
\newblock \showarticletitle{Adaptive Time Slotted Channel Hopping for Wireless
  Sensor Networks}. In \bibinfo{booktitle}{\emph{2012 4th Computer Science and
  Electronic Engineering Conference (CEEC '12)}}. \bibinfo{pages}{29--34}.
\newblock
\urldef\tempurl%
\url{https://doi.org/10.1109/CEEC.2012.6375374}
\showDOI{\tempurl}


\bibitem[Elsts et~al\mbox{.}(2017)]%
        {active_ch_5}
\bibfield{author}{\bibinfo{person}{A. Elsts}, \bibinfo{person}{X. Fafoutis},
  \bibinfo{person}{R. Piechocki}, {and} \bibinfo{person}{I. Craddock}.}
  \bibinfo{year}{2017}\natexlab{}.
\newblock \showarticletitle{Adaptive Channel Selection in IEEE 802.15.4 TSCH
  Networks}. In \bibinfo{booktitle}{\emph{1st Global Internet of Things Summit
  (GIoTS '17)}}. \bibinfo{pages}{1--6}.
\newblock
\urldef\tempurl%
\url{https://doi.org/10.1109/GIOTS.2017.8016246}
\showDOI{\tempurl}


\bibitem[Gabillon et~al\mbox{.}(2012)]%
        {Best_Arm_Identification}
\bibfield{author}{\bibinfo{person}{V. Gabillon}, \bibinfo{person}{M.
  Ghavamzadeh}, {and} \bibinfo{person}{A. Lazaric}.}
  \bibinfo{year}{2012}\natexlab{}.
\newblock \showarticletitle{Best Arm Identification: A Unified Approach to
  Fixed Budget and Fixed Confidence}. In \bibinfo{booktitle}{\emph{Proceedings
  of the 25th International Conference on Neural Information Processing Systems
  (NIPS '12)}}, Vol.~\bibinfo{volume}{2}. \bibinfo{pages}{3212–3220}.
\newblock


\bibitem[Gomes et~al\mbox{.}(2017)]%
        {active_ch_3}
\bibfield{author}{\bibinfo{person}{P.~H. Gomes}, \bibinfo{person}{T. Watteyne},
  {and} \bibinfo{person}{B. Krishnamachari}.} \bibinfo{year}{2017}\natexlab{}.
\newblock \showarticletitle{MABO-TSCH: Multi-hop And Blacklist-based Optimized
  Time Synchronized Channel Hopping}.
\newblock \bibinfo{journal}{\emph{Transactions on Emerging Telecommunications
  Technologies}}  \bibinfo{volume}{29} (\bibinfo{date}{08}
  \bibinfo{year}{2017}), \bibinfo{pages}{e3223}.
\newblock
\urldef\tempurl%
\url{https://doi.org/10.1002/ett.3223}
\showDOI{\tempurl}


\bibitem[Grant(2016)]%
        {LPWAN_NB_IoT}
\bibfield{author}{\bibinfo{person}{S. Grant}.} \bibinfo{year}{2016}\natexlab{}.
\newblock \bibinfo{title}{3GPP Low Power Wide Area Technologies}.
\newblock \bibinfo{howpublished}{GSMA}.
\newblock
\urldef\tempurl%
\url{https://www.gsma.com/iot/resources/3gpp-low-power-wide-area-technologies-white-paper/}
\showURL{%
\tempurl}


\bibitem[H{\"a}nninen et~al\mbox{.}(2011)]%
        {active_ch_1}
\bibfield{author}{\bibinfo{person}{M. H{\"a}nninen}, \bibinfo{person}{J.
  Suhonen}, \bibinfo{person}{T.~D. H{\"a}m{\"a}l{\"a}inen}, {and}
  \bibinfo{person}{M. H{\"a}nnik{\"a}inen}.} \bibinfo{year}{2011}\natexlab{}.
\newblock \showarticletitle{Link Quality-Based Channel Selection for Resource
  Constrained WSNs}.
\newblock \bibinfo{journal}{\emph{Lecture Notes in Computer Science}}
  \bibinfo{volume}{6646} (\bibinfo{year}{2011}), \bibinfo{pages}{254--263}.
\newblock
\urldef\tempurl%
\url{https://doi.org/10.1007/978-3-642-20754-9_26}
\showDOI{\tempurl}


\bibitem[{Hasegawa} et~al\mbox{.}(2022)]%
        {active_ch_8}
\bibfield{author}{\bibinfo{person}{S. {Hasegawa}}, \bibinfo{person}{R.
  {Kitagawa}}, \bibinfo{person}{A. {Li}}, \bibinfo{person}{S.~J. {Kim}},
  \bibinfo{person}{Y. {Watanabe}}, \bibinfo{person}{Y. {Shoji}}, {and}
  \bibinfo{person}{M. {Hasegawa}}.} \bibinfo{year}{2022}\natexlab{}.
\newblock \showarticletitle{Multi-Armed-Bandit Based Channel Selection
  Algorithm for Massive Heterogeneous Internet of Things Networks}.
\newblock \bibinfo{journal}{\emph{Applied Sciences}} \bibinfo{volume}{12},
  \bibinfo{number}{15} (\bibinfo{year}{2022}), \bibinfo{pages}{7424}.
\newblock
\urldef\tempurl%
\url{https://doi.org/10.3390/app12157424}
\showDOI{\tempurl}


\bibitem[Ibáñez et~al\mbox{.}(2017)]%
        {LoRa_Power}
\bibfield{author}{\bibinfo{person}{L.~C. Ibáñez}, \bibinfo{person}{B.~M.
  Masnou}, \bibinfo{person}{R.~V. Ferré}, {and} \bibinfo{person}{C. Gomez}.}
  \bibinfo{year}{2017}\natexlab{}.
\newblock \showarticletitle{Modeling the energy performance of LoRaWAN}.
\newblock \bibinfo{journal}{\emph{Sensors}} \bibinfo{volume}{17},
  \bibinfo{number}{10} (\bibinfo{date}{Oct} \bibinfo{year}{2017}),
  \bibinfo{pages}{2364}.
\newblock
\urldef\tempurl%
\url{https://doi.org/10.3390/s17102364}
\showDOI{\tempurl}


\bibitem[Katehakis and Veinott(1987)]%
        {MAB}
\bibfield{author}{\bibinfo{person}{M.~N. Katehakis} {and}
  \bibinfo{person}{A.~F. Veinott}.} \bibinfo{year}{1987}\natexlab{}.
\newblock \showarticletitle{The Multi-Armed Bandit Problem: Decomposition and
  Computation}.
\newblock \bibinfo{journal}{\emph{Mathematics of Operations Research}}
  \bibinfo{volume}{12} (\bibinfo{year}{1987}), \bibinfo{pages}{262--268}.
\newblock
\urldef\tempurl%
\url{https://doi.org/10.1287/moor.12.2.262}
\showDOI{\tempurl}


\bibitem[Kotsiou et~al\mbox{.}(2017)]%
        {active_ch_4}
\bibfield{author}{\bibinfo{person}{V. Kotsiou}, \bibinfo{person}{G.~Z.
  Papadopoulos}, \bibinfo{person}{P. Chatzimisios}, {and} \bibinfo{person}{F.
  Theoleyre}.} \bibinfo{year}{2017}\natexlab{}.
\newblock \showarticletitle{LABeL: Link-Based Adaptive BLacklisting Technique
  for 6TiSCH Wireless Industrial Networks}. In
  \bibinfo{booktitle}{\emph{Proceedings of the 20th ACM International
  Conference on Modelling, Analysis and Simulation of Wireless and Mobile
  Systems (MSWiM '17)}}. \bibinfo{pages}{25–33}.
\newblock
\urldef\tempurl%
\url{https://doi.org/10.1145/3127540.3127541}
\showDOI{\tempurl}


\bibitem[Li et~al\mbox{.}(2012)]%
        {active_ch_2}
\bibfield{author}{\bibinfo{person}{J. Li}, \bibinfo{person}{W. Zeng}, {and}
  \bibinfo{person}{A. Arora}.} \bibinfo{year}{2012}\natexlab{}.
\newblock \showarticletitle{Chameleon: On the Energy Efficiency of Exploiting
  Multiple Frequencies in Wireless Sensor Networks}. In
  \bibinfo{booktitle}{\emph{Broadband Communications, Networks, and Systems.
  7th International ICST Conference, (BROADNETS '10)}}.
\newblock
\urldef\tempurl%
\url{https://doi.org/10.1007/978-3-642-30376-0_10}
\showDOI{\tempurl}


\bibitem[{Romero} et~al\mbox{.}(2021)]%
        {active_ch_6}
\bibfield{author}{\bibinfo{person}{N.~C. {Romero}}, \bibinfo{person}{J.~N.
  {Ortiz}}, \bibinfo{person}{P. {Muñoz}}, {and} \bibinfo{person}{P.
  {Ameigeiras}}.} \bibinfo{year}{2021}\natexlab{}.
\newblock \showarticletitle{Collision Avoidance Resource Allocation for
  LoRaWAN}.
\newblock \bibinfo{journal}{\emph{Sensors}} \bibinfo{volume}{21},
  \bibinfo{number}{4} (\bibinfo{date}{Feb} \bibinfo{year}{2021}),
  \bibinfo{pages}{1218}.
\newblock
\urldef\tempurl%
\url{https://doi.org/10.3390/s21041218}
\showDOI{\tempurl}


\bibitem[Sahai et~al\mbox{.}(2004)]%
        {passive_ch_1}
\bibfield{author}{\bibinfo{person}{A. Sahai}, \bibinfo{person}{N. Hoven}, {and}
  \bibinfo{person}{R. Tandra}.} \bibinfo{year}{2004}\natexlab{}.
\newblock \showarticletitle{Some Fundamental Limits on Cognitive Radio}. In
  \bibinfo{booktitle}{\emph{2004 42th Annual Allerton Conference on
  Communication, Control, and Computing (Allerton '04)}}.
  \bibinfo{pages}{1662--1671}.
\newblock


\bibitem[Semtech({[n.\,d.]})]%
        {LPWAN_LoRa}
\bibfield{author}{\bibinfo{person}{Semtech}.}
  \bibinfo{year}{[n.\,d.]}\natexlab{}.
\newblock \bibinfo{booktitle}{\emph{LoRa(PHY)}}.
\newblock
\urldef\tempurl%
\url{https://www.semtech.com/lora/what-is-lora}
\showURL{%
\tempurl}


\bibitem[Semtech(2015)]%
        {sx1276}
\bibfield{author}{\bibinfo{person}{Semtech}.} \bibinfo{year}{2015}\natexlab{}.
\newblock \bibinfo{booktitle}{\emph{137 MHz to 1020 MHz Low Power Long Range
  Transceiver}}.
\newblock
\newblock
\shownote{Rev. 4}.


\bibitem[Sigfox({[n.\,d.]})]%
        {LPWAN_Sigfox}
\bibfield{author}{\bibinfo{person}{Sigfox}.}
  \bibinfo{year}{[n.\,d.]}\natexlab{}.
\newblock \bibinfo{booktitle}{\emph{Sigfox 0G Technology}}.
\newblock
\urldef\tempurl%
\url{https://www.sigfox.com}
\showURL{%
\tempurl}


\bibitem[Tang et~al\mbox{.}(2011)]%
        {passive_ch_2}
\bibfield{author}{\bibinfo{person}{L. Tang}, \bibinfo{person}{Y. Sun},
  \bibinfo{person}{O. Gurewitz}, {and} \bibinfo{person}{D. Johnson}.}
  \bibinfo{year}{2011}\natexlab{}.
\newblock \showarticletitle{EM-MAC: A Dynamic Multichannel Energy-Efficient MAC
  Protocol for Wireless Sensor Networks}. In
  \bibinfo{booktitle}{\emph{Proceedings of the 12th ACM International Symposium
  on Mobile Ad Hoc Networking and Computing (MobiHoc '11)}}. Article
  \bibinfo{articleno}{23}, \bibinfo{numpages}{11}~pages.
\newblock
\urldef\tempurl%
\url{https://doi.org/10.1145/2107502.2107533}
\showDOI{\tempurl}


\bibitem[Tavakoli et~al\mbox{.}(2018)]%
        {passive_ch_5}
\bibfield{author}{\bibinfo{person}{R. Tavakoli}, \bibinfo{person}{M. Nabi},
  \bibinfo{person}{T. Basten}, {and} \bibinfo{person}{K. Goossens}.}
  \bibinfo{year}{2018}\natexlab{}.
\newblock \showarticletitle{Dependable Interference-Aware Time-Slotted Channel
  Hopping for Wireless Sensor Networks}.
\newblock \bibinfo{journal}{\emph{ACM Transactions on Sensor Networks}}
  \bibinfo{volume}{14}, \bibinfo{number}{1}, Article \bibinfo{articleno}{3}
  (\bibinfo{year}{2018}), \bibinfo{numpages}{35}~pages.
\newblock
\urldef\tempurl%
\url{https://doi.org/10.1145/3158231}
\showURL{%
\tempurl}


\bibitem[{Thompson}(1933)]%
        {Thompson1933}
\bibfield{author}{\bibinfo{person}{W.~R. {Thompson}}.}
  \bibinfo{year}{1933}\natexlab{}.
\newblock \showarticletitle{On The Likelihood That One Unknown Probability
  Exceeds Another in View of The Evidence of Two Samples}.
\newblock \bibinfo{journal}{\emph{Biometrika}}  \bibinfo{volume}{25}
  (\bibinfo{year}{1933}), \bibinfo{pages}{285--294}.
\newblock


\bibitem[{Thompson}(1935)]%
        {Thompson1935}
\bibfield{author}{\bibinfo{person}{W.~R. {Thompson}}.}
  \bibinfo{year}{1935}\natexlab{}.
\newblock \showarticletitle{On the Theory of Apportionment}.
\newblock \bibinfo{journal}{\emph{American Journal of Mathematics}}
  \bibinfo{volume}{57} (\bibinfo{year}{1935}), \bibinfo{pages}{450}.
\newblock


\bibitem[{Wi-Fi Alliance}({[n.\,d.]})]%
        {LPWAN_HaLow}
\bibfield{author}{\bibinfo{person}{{Wi-Fi Alliance}}.}
  \bibinfo{year}{[n.\,d.]}\natexlab{}.
\newblock \bibinfo{booktitle}{\emph{Wi-Fi Certified HaLow}}.
\newblock
\urldef\tempurl%
\url{https://www.wi-fi.org/discover-wi-fi/wi-fi-certified-halow}
\showURL{%
\tempurl}


\bibitem[Yun et~al\mbox{.}(2022)]%
        {MKII}
\bibfield{author}{\bibinfo{person}{J. Yun}, \bibinfo{person}{S. Srivastava},
  \bibinfo{person}{D. Roy}, \bibinfo{person}{N. Stohs}, \bibinfo{person}{C.
  Mydlarz}, \bibinfo{person}{M. Salman}, \bibinfo{person}{B. Steers},
  \bibinfo{person}{J.~P. Bello}, {and} \bibinfo{person}{A. Arora}.}
  \bibinfo{year}{2022}\natexlab{}.
\newblock \showarticletitle{Infrastructure-free, Deep Learned Urban Noise
  Monitoring at ~100mW}. In \bibinfo{booktitle}{\emph{2022 ACM/IEEE 13th
  International Conference on Cyber-Physical Systems (ICCPS '22)}}.
  \bibinfo{pages}{56--67}.
\newblock
\urldef\tempurl%
\url{https://doi.org/10.1109/ICCPS54341.2022.00012}
\showDOI{\tempurl}


\end{thebibliography}

%\printbibliography

\end{document}